\documentclass[aps,prl,reprint,twocolumn,showpacs,superscriptaddress,longbibliography]{revtex4-2} 

\usepackage{graphicx}
\usepackage{dcolumn}
\usepackage{bm}

\usepackage{color}
\usepackage{xcolor}

\begin{document}

\title{Epithelial Tissue Growth Dynamics: Universal or Not?}

\author{Mahmood Mazarei}
\affiliation{Department of Mathematics,
  The University of Western Ontario, 1151 Richmond Street, London,
  Ontario, Canada  N6A\,5B7}
\affiliation{The Centre for Advanced Materials and Biomaterials Research,
  The University of Western Ontario, 1151 Richmond Street,
  London, Ontario, Canada N6A\,3K7}

\author{Jan {\AA}str{\"o}m}
\affiliation{CSC Scientific Computing Ltd, K{\"a}gelstranden 14, 02150
  Esbo, Finland}

\author{Jan Westerholm}
\affiliation{Faculty of Science and Engineering, {\AA}bo Akademi
University, Vattenborgsv\"agen 3, FI-20500, {\AA}bo, Finland
}

\author{Mikko Karttunen}
\affiliation{The Centre for Advanced Materials and Biomaterials Research,
  The University of Western Ontario, 1151 Richmond Street,
London, Ontario, Canada N6A\,3K7}
\affiliation{Department of Chemistry,  The University of Western Ontario, 1151 Richmond Street, London,
  Ontario, Canada N6A\,5B7
}
\affiliation{Department of Physics and Astronomy,
  The University of Western Ontario, 1151 Richmond Street, London,
  Ontario, Canada  N6A\,3K7}

\date{\today}

\begin{abstract}
Universality of interfacial roughness 
in growing epithelial tissue has remained a controversial issue. Kardar-Parisi-Zhang (KPZ) and Molecular Beam Epitaxy (MBE) universality classes have been reported among other behaviors including total lack of universality.
Here, we utilize a kinetic division model for deformable cells to investigate cell-colony scaling. With seemingly minor model changes, it can reproduce both KPZ- and MBE-like scaling in configurations that mimic the respective experiments. This result neutralizes the apparent scaling controversy. 
It can be speculated that this diversity in growth behavior 
is  beneficial for efficient evolution and versatile growth dynamics.
\end{abstract}
\maketitle

Growth of biological matter, e.g., tumor invasion, depends on complex processes such as the mechanism(s) of proliferation, the physical properties of the microenvironment, and cellular migration that can be dominated either by single cell or collective motion that depend on  intercellular interactions and intracellular regulation~\cite{sengupta2021principles}.

Characterization of interfacial growth of cellular colonies is important for understanding the factors that
control growth and how they manifest themselves in the kinetics and morphology of cell aggregation. 
Numerous  experimental and computational studies have investigated the effects of biochemical regulation and mechanical factors such as cell-to-cell adhesion and friction, and cell division~\cite{Lecuit2007-fk,costa2015universal,li2021role,khain2021dynamics,radszuweit2009comparing,Bru2003,Bru1998,Bru2005,Huergo2010,Huergo2011}. Typically,  growth is characterized by scaling analysis  which  identifies the underlying mechanisms of growth dynamics by critical exponents. 

Self-affinity of the interface width, $w(l,t)$, obeys the Family-Vicsek scaling relation~\cite{Family1985,Family1990a}
\begin{equation}
\label{eq:1}
w(l,t) \sim t^{\beta} F(lt^{- \frac{1}{z}}),
\end{equation}
where the exponent $z$ describes the scaling relation between the critical time and length scales, and can be obtained from the scaling relation
%
$z = \frac{\alpha}{\beta}$,
%
where the exponent $\alpha$ characterizes the roughness of the interface. The exponent $\beta$
is obtained through the scaling function $F(u) = F(lt^{- \frac{1}{z}})$ which has the following properties:
There is a crossover at $u=l_\mathrm{*}$. For $u\ll l_\mathrm{*}$ the scaling function increases as a power law, $F(u)=u^{\beta}$, where $\beta$ is the growth exponent and characterizes the time-dependent dynamics surface roughening. For $u\gg l_\mathrm{*}$ the width saturates, and $F(u)$ becomes a constant~\cite{Barabasi1995}. 
With these three critical exponents, interfacial growth is often classified into different dynamic universality classes. 

The KPZ equation was the first nonlinear continuum equation used to study surface growth. It is described by the stochastic differential equation~\cite{Kardar1986}
\begin{equation}
\label{eq:3}
\partial_{t}h(x,t) = -\lambda \left[\partial_{x} h(x,t)\right]^{2} + \nu \partial_{x}^{2} h(x,t) + \xi(x,t), 
\end{equation} 
where the height ($h(x,t)$) depends on position and time, and $\lambda, \nu$ and $D$ are physical constants.  
The first term on the RHS reflects growth that occurs locally normal to the interface and
renders the KPZ equation nonlinear. The second term  smooths the interface by surface tension $\nu$, and the last term, $\xi(x,t)$, is Gaussian noise given
by $\langle \xi(x,t) \rangle=0$ and $\langle \xi(s,x) \xi(t,y)\rangle \! = \! 2 D \delta(s-t) \delta(x-y)$. The KPZ  universality class is characterized by the exponents $\alpha^\mathrm{KPZ} \! = \! \frac{1}{2}$, 
$\beta^\mathrm{KPZ} \! = \! \frac{1}{3}$, and $z^\mathrm{KPZ} \! = \! \frac{3}{2}$~\cite{Kardar1986}. 

Mathematically, surface tension and lateral growth determine the asymptotic scaling of the KPZ equation. In some growth processes, however, surface diffusion controls the scaling behavior, and the growth process is described by the MBE model~\cite{Kessler1992,Sarma1994}
\begin{equation}
\label{eq:4}
\partial_{t}h(x,t) = - K \partial_{x}^{4} h(x,t) + F + \xi(x,t), 
\end{equation} 
where $K$ is the surface diffusion coefficient, $F$ is the growth rate, and $\xi(x,t)$ is 
Gaussian white noise 
as in the KPZ equation. The roughness exponents for the MBE universality class for a one dimensional interface are $\alpha^\mathrm{MBE} \!= \! \frac{3}{2}$, $\beta^\mathrm{MBE} \!= \! \frac{3}{8}$, and $ z^\mathrm{MBE} \! = \! 4.0$.

Br\'u \textit{et al.} studied  cellular growth using cells from 15 different \textit{in vitro} cell lines and 16 \textit{in vivo} types of tumor cells obtained from patients~\cite{Bru2003,Bru1998}. They determined the growth to belong to the MBE universality class  in all cases  with exponents $\alpha \! = \!  1.5 \! \pm \! 0.15$, $\beta \! = \! 0.38 \! \pm \! 0.07$, and $z \! = \! 4.0 \! \pm \! 0.5$ thus suggesting universal growth dynamics for cells.  This conclusion was strongly criticized by Buceta and Galeano~\cite{Buceta2005} who dismissed  universality of tumor growth dynamics stating serious flaws in Br\'u \textit{et al.}'s scaling analysis. In their rebuttal Br\'u \textit{et al.}~\cite{Bru2005} restated their conclusions and wrote  \textit{"the  characteristics of MBE dynamics discussed in Br\'u et al. (2003,1998) have not only been rigorously demonstrated but have served  as the  basis for  a successful antitumor therapy currently   under   development."} A recent study of growth of different brain tumors \textit{in vivo} using fractal and scaling analysis shows similarities with some of the results of Br\'u \textit{et al.}~\cite{hoyos2018geometrical}. 

In contrast to MBE-like dynamics,  Huergo \textit{et al.}~\cite{Huergo2010,Huergo2011,Huergo2012} reported 
KPZ scaling for both linearly and radially spreading interfaces of HeLa (cervix cancer) 
and Vero cell colonies. Galeano \textit{et al.} studied the development of 
plant cell species \textit{Brassica oleracea} and \textit{B. rapa} under various growing conditions and 
obtained  $\alpha \! = \! 0.86 \pm 0.4$, and $z \! = \! 5.0$~\cite{galeano2003dynamical}. Santalla \textit{et al.}~\cite{santalla2018nonuniversality} grew colonies of \textit{Bacillus subtilis} and \textit{Escherichia coli}
using a high agar concentration regime with various nutrients and discovered branching interfaces with 
exponents  $\beta \! = \! 0.5$ and $\alpha \! = \! 0.75$ that are inconsistent with both MBE and KPZ.

Substrate disorder  can also influence growth dynamics. Vicsek \textit{et al.}~\cite{Vicsek1990} studied the growth of \textit{E. coli} and \textit{B. subtilis} colonies and found the roughness exponent $\alpha \! = \! 0.78 \pm 0.07$  which is inconsistent with both the  KPZ and MBE models as well as with the quenched KPZ (qKPZ) model~\cite{Barabasi1995} that includes disorder. It has also been demonstrated that the  behavior of bacterial colonies in the medium-to-high nutrient concentration regime  can be  very rich due to the appearance of quenched disorder in growth patterns~\cite{Bonachela2011}.  In that context, Huergo \textit{et al.} also examined the 2D growth dynamics of quasilinear Vero cell colony fronts in a methylcellulose-containing culture medium. Their scaling analysis yielded $\alpha \!=\! 0.63 \pm 0.04$, $\beta \! = \! 0.75 \pm 0.05$, and $z\! =\! 0.84 \pm 0.05$, suggesting qKPZ dynamics~\cite{Huergo2014}.

On the computational and theoretical side, Santalla and Ferreira~\cite{santalla2018eden} used an off-lattice Eden model modified to account for nutrient diffusion. Under scarce nutrient supply, they observed initially a KPZ regime that transitioned via a qKPZ transient to unstable growth. Block \textit{et al.} studied the growth 
of 2D cellular monolayers for a class of cellular automaton models. Their  results suggest KPZ dynamics over a wide range of parameters and different cell migration dynamics~\cite{Block2007} contradicting the MBE dynamics reported by Br\'u \textit{et al.}~\cite{Bru2003}. Another contradiction was reported by Azimzade \textit{et al.} who developed a tumor growth model based on the nonlinear Fisher-Kolmogorov-Petrovsky-Piskunov equation, a reaction-diffusion equation, to investigate the impact of the cellular environment and spatial correlations  on
tumor invasion~\cite{azimzade2019effect}. They concluded that kinetic growth models, such as KPZ, cannot characterise tumor invasion fronts, and that the structure of the tumor interface depends intimately on the initial conditions~\cite{azimzade2019effect}.

A large number of different models has been used to describe cellular growth~\cite{Buttenschon2020-wh}.
Here, we use the \textit{CellSim3D} off-lattice growth model and simulator to study epithelial tissue growth~\cite{Madhikar2018,Madhikar2020}. In this model, cells can migrate, deform, divide, and interact with each other and their environment mechanically via adhesion and friction.  Its 2D version has been shown to produce, e.g., cell-cell force distributions, force dipoles, spontaneous orientation of cells in the direction of highest stiffness and cellular migration in agreement with experiments~\cite{Madhikar2021-ko}.  \textit{CellSim3D} software leverages graphics processor units, enabling simulations of systems $> \! \! 100,000$ cells easily.
The analysis below uses averages over 10 independent simulations. Details, parameters and animations demonstrating the model are provided as Supplemental Material. In brief, in \textit{CellSim3D}, epithelial tissues can be modeled as quasi-2D systems of 3D cells confined into a plane, corresponding to the experimental confinement of cells between two plates; the bottom plate models basal tissue and the top plate prevents excessive buckling.

\begin{figure}
\resizebox{0.48\columnwidth}{!}{ \includegraphics{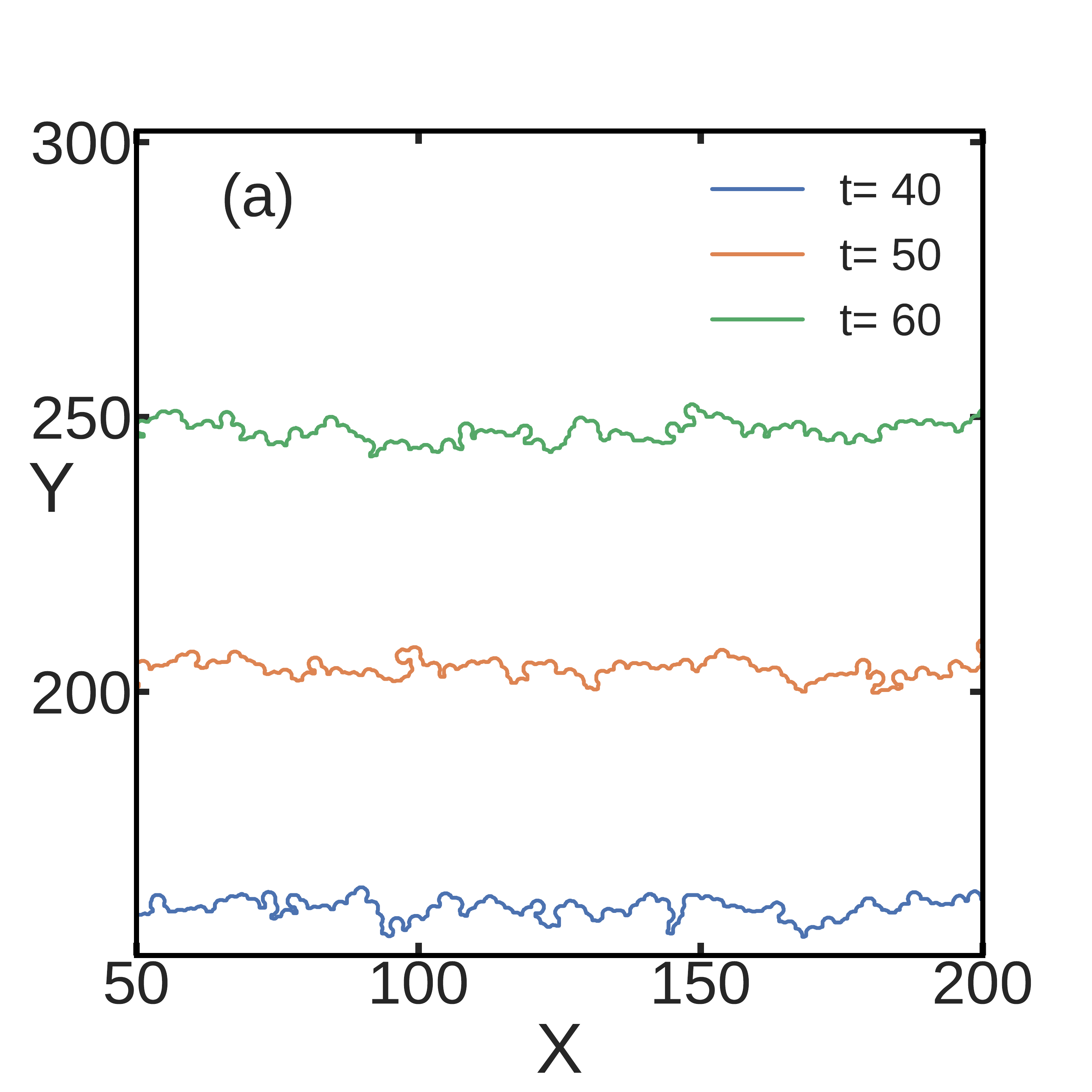} }
\resizebox{0.48\columnwidth}{!}{ \includegraphics{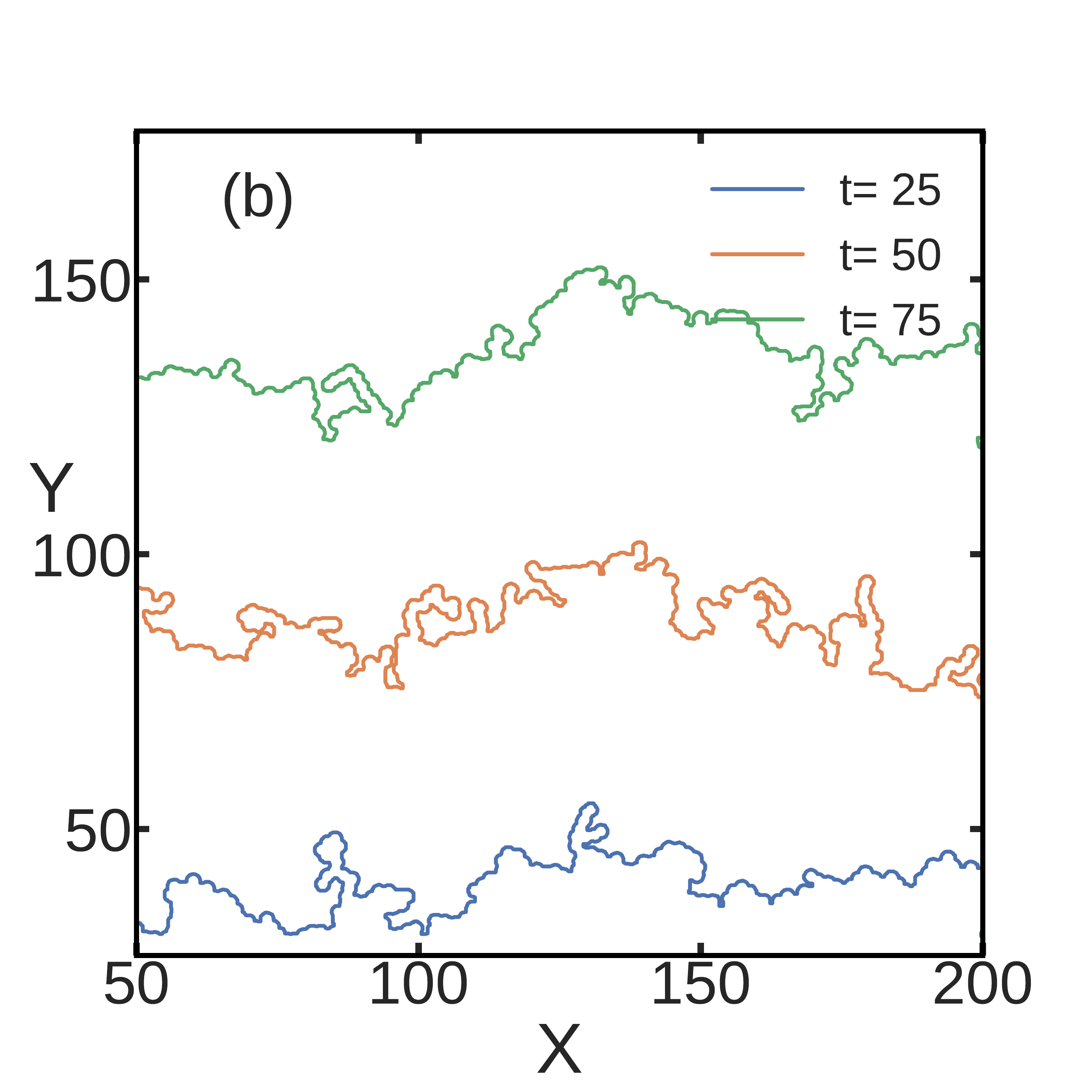} }
\resizebox{0.48\columnwidth}{!}{ \includegraphics{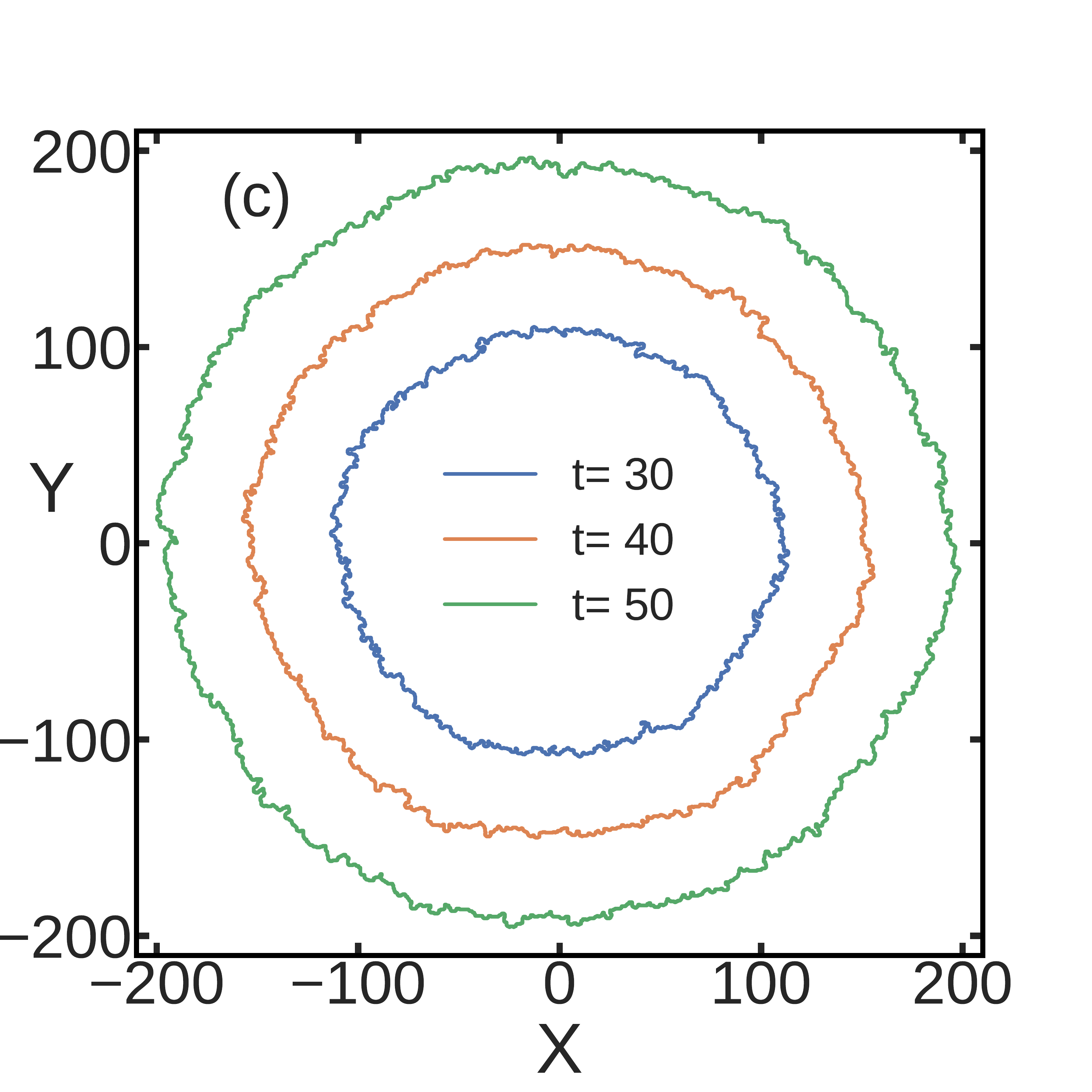} }
\resizebox{0.48\columnwidth}{!}{ \includegraphics{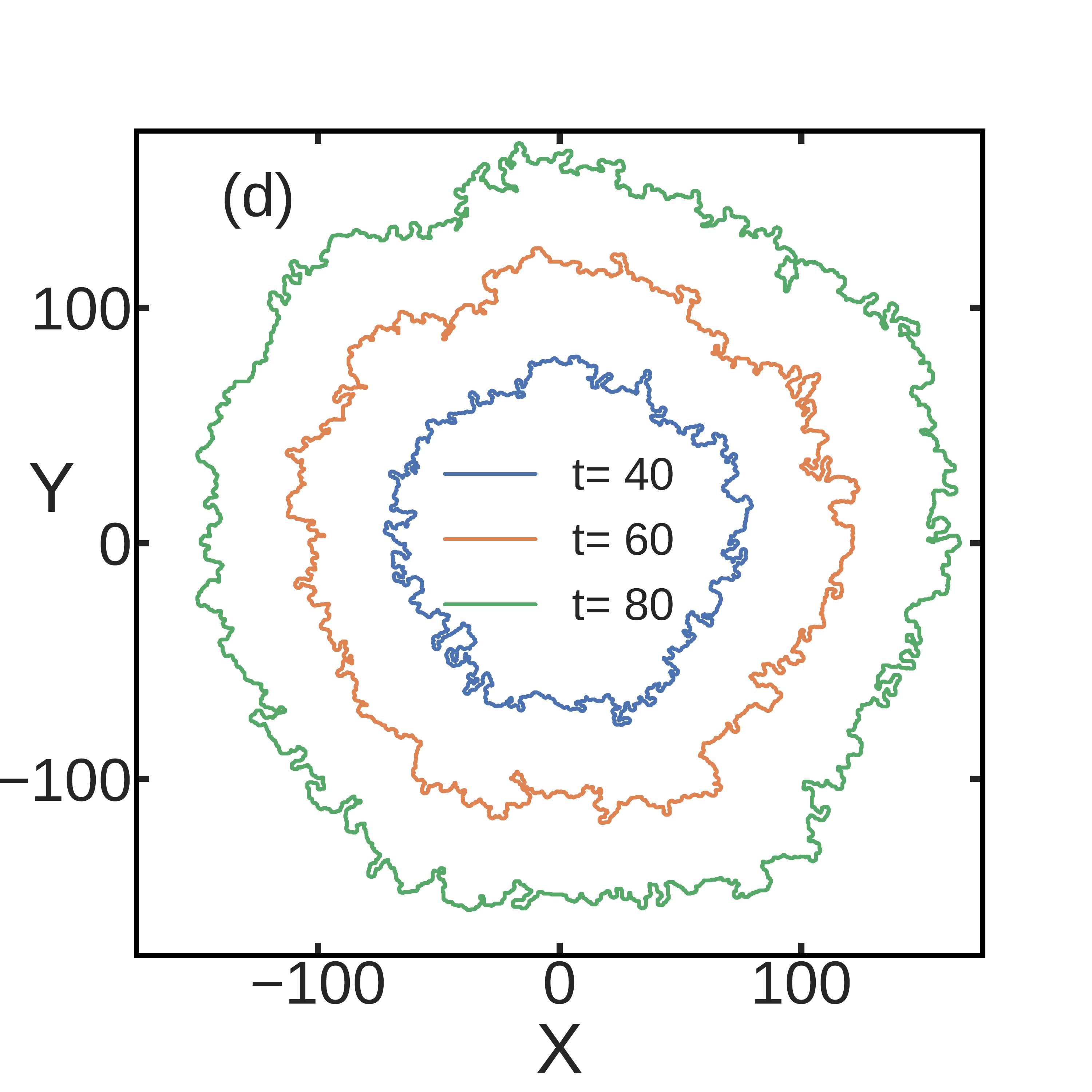} }
\caption{Interface evolutions of cell colonies starting from a horizontal line and a radially expanding interface.
(a) Line configuration at weak ($10$) and (b) strong cell-cell adhesion ($2000$). (c) Radially expanding interface with at weak ($10$) and (d) strong adhesion ($2000$). All interfaces have overhangs. Scaling analysis was done using overhang-corrected interfaces~\cite{Barabasi1995}. For the units,
see Table~S1.}
\label{fig:fig101}
\end{figure}

The  interface width is defined as the standard deviation of the height over a length scale $l$ at time $t$ as~\cite{Barabasi1995}
\begin{equation}
\label{eq:5}
w(l,t) = \bigg \lbrace \frac{1}{N} \sum_{i=1}^{N} [h_{i}(t) - \langle h_{i} \rangle_{l}]^{2} \bigg \rbrace _{L}^{\frac{1}{2}},
\end{equation} 
where $L$ is the contour length, which increases with time as $L=2\pi\langle h(t)\rangle$ for radially expanding fronts and is constant for linear fronts. For radially expanding fronts,  $h_\mathrm{i}(t)$ is the distance from the center of mass to the point $i$ of the interface at time $t$, $\langle h_\mathrm{i} \rangle_{l}$ is the local average of the subsets of arc length $l$, and $\lbrace \cdot \rbrace_{L}$ is the overall average. 
We complement the universality class analysis by an examination of the structure factor, 
\begin{equation}
\label{eq:8}
S(k,t) = \langle \hat{h}(k,t) \hat{h}(-k,t) \rangle ,
\end{equation}
where $k$ is the wavenumber, and $\hat{h}(k,t)$ is the Fourier transform of the interface profile $h(x,t)$ \cite{Barabasi1995}. The advantage of this method over the real space is that only long-wavelength modes contribute to its scaling. Hence, it is less affected by finite-size effects. This method provides the global roughness exponent $\alpha$ and the dynamic exponent $z$ via the Family-Vicsek scaling form 
for $S(k,t)$,
\begin{equation}
\label{eq:9}
S(k,t) = k^{-2\alpha + 1} s(kt^{\frac{1}{z}}), \mathrm{\,\,where}
\end{equation}
\begin{equation}
\label{eq:10}
s(u=kt^{\frac{1}{z}}) = 
 \left\{ \begin{array}{ll}
         \mathrm{const} & \mbox{for $u\gg1$};\\
         u^{-2\alpha + 1} & \mbox{for $u\ll1$}.\end{array} \right.
\end{equation}
At $u \! = \!1$ there is a crossover, for $u \! \gg \! 1$ the curves measured at different 
times collapse 
and for $u \! \ll \! 1$ they split. 

\begin{figure}
\resizebox{0.49\columnwidth}{!}{ \includegraphics{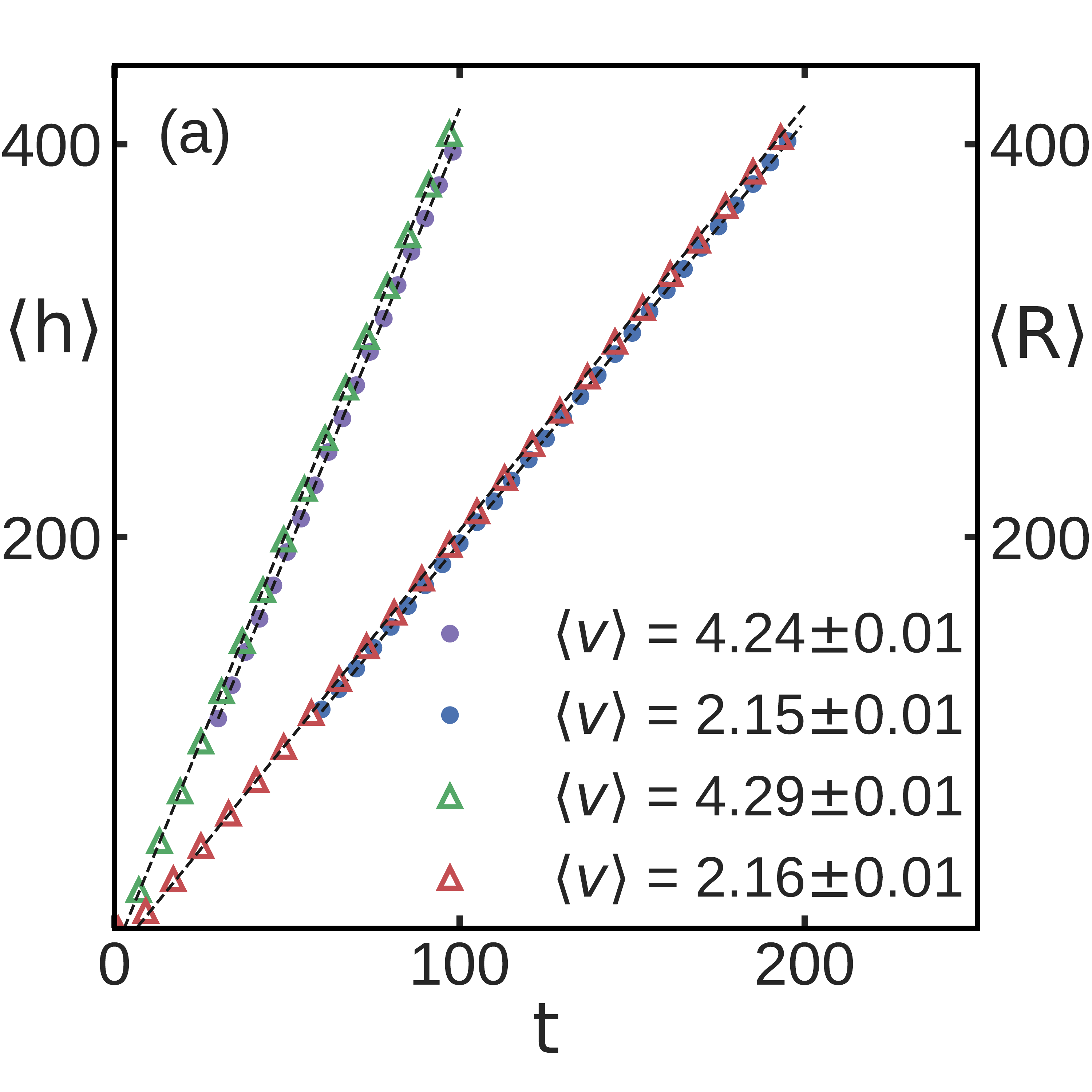} }
\resizebox{0.49\columnwidth}{!}{ \includegraphics{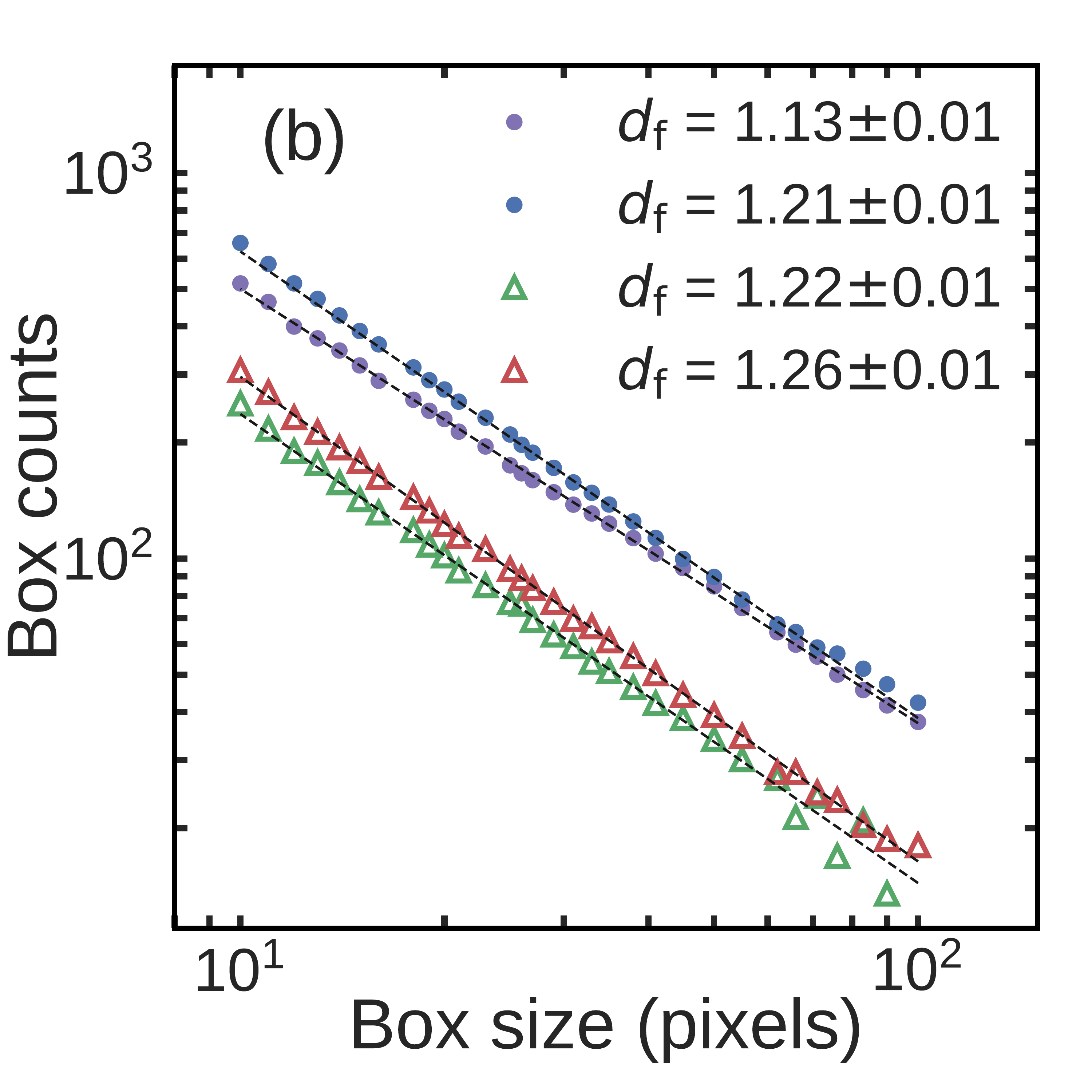} }
\caption{ (a) 
Velocity ($\langle v \rangle$) of the interface determined from the mean colony radius $\langle R \rangle$ and the mean interface height $\langle h \rangle$ vs. time for different cell-cell adhesion strengths.  Radially expanding interface:  (purple circles) at weak ($10$)  and (blue circles) strong adhesion ($2,000$).  Line configuration: (green triangles) at weak and (red triangles) at strong adhesion.  (b) The fractal dimension ($d_\mathrm{f}$) determined by plotting box counts vs. box size. For radially expanding interface: (purple circles) at weak ($10$)  and (blue circles)  strong adhesion ($2,000$).  For line configuration: (green triangles) at weak and (red triangles) strong adhesion. For units, see Table~S1.
}
\label{fig:fig102}
\end{figure}

\begin{figure}
\resizebox{0.49\columnwidth}{!}{ \includegraphics{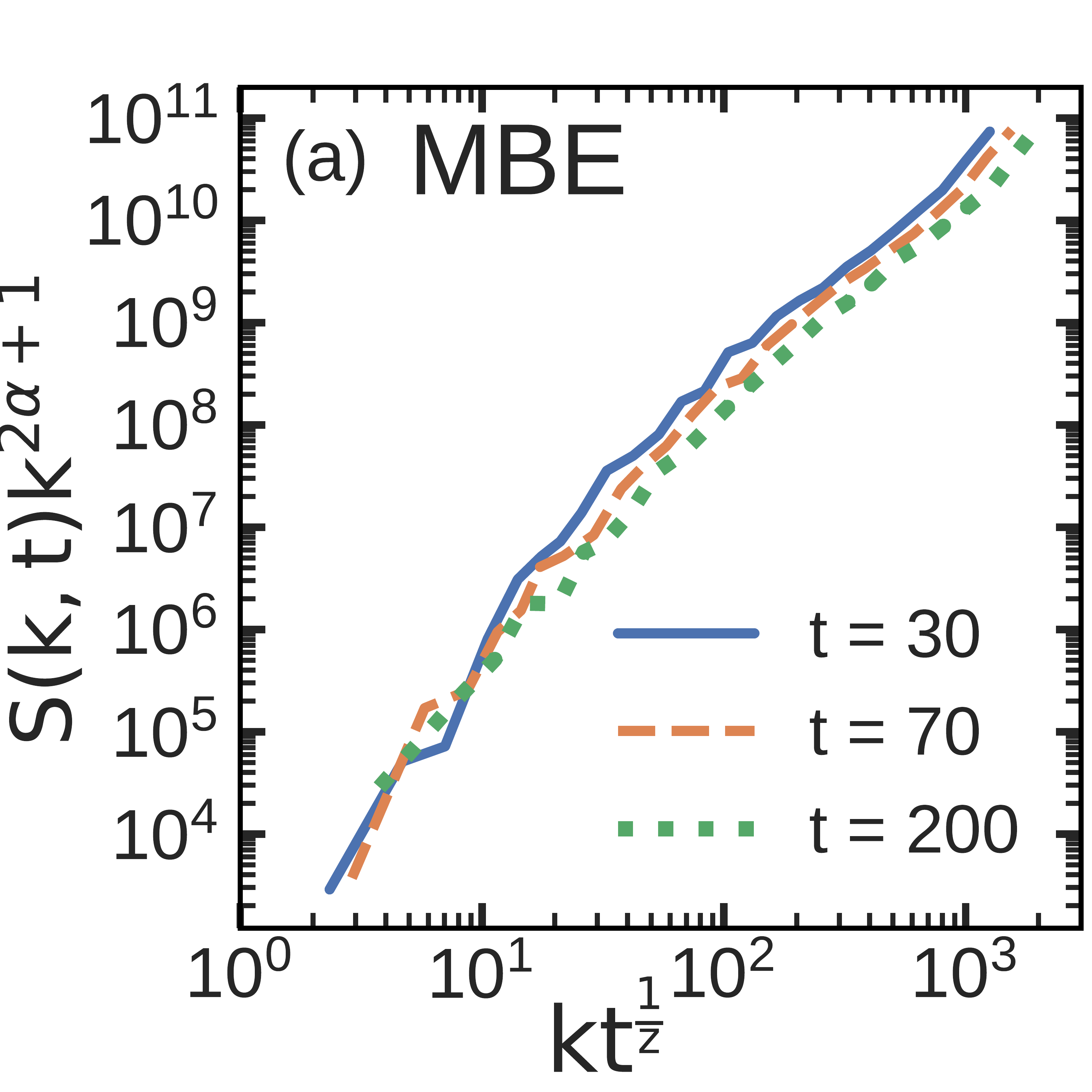} }
\resizebox{0.49\columnwidth}{!}{ \includegraphics{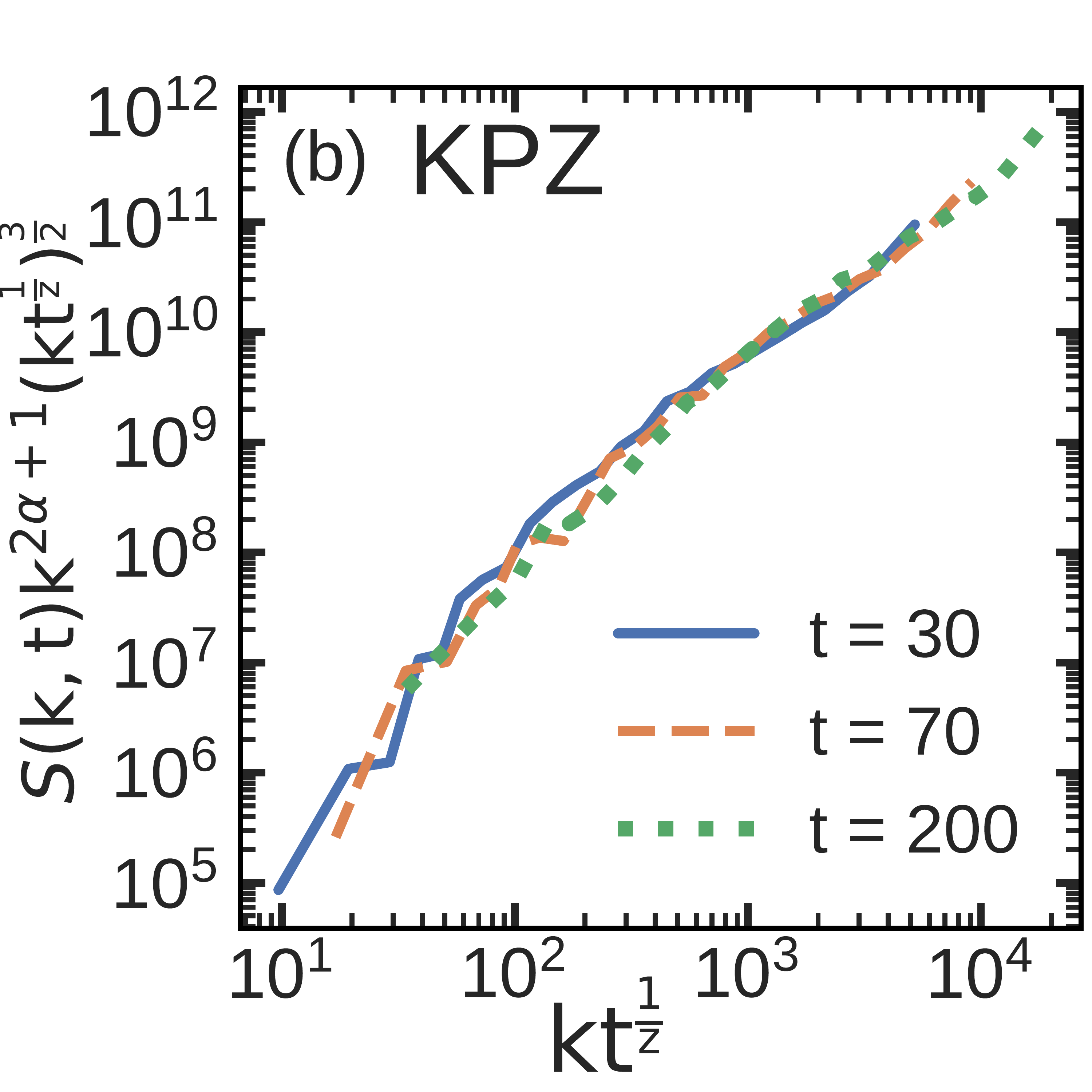} }
\resizebox{0.49\columnwidth}{!}{ \includegraphics{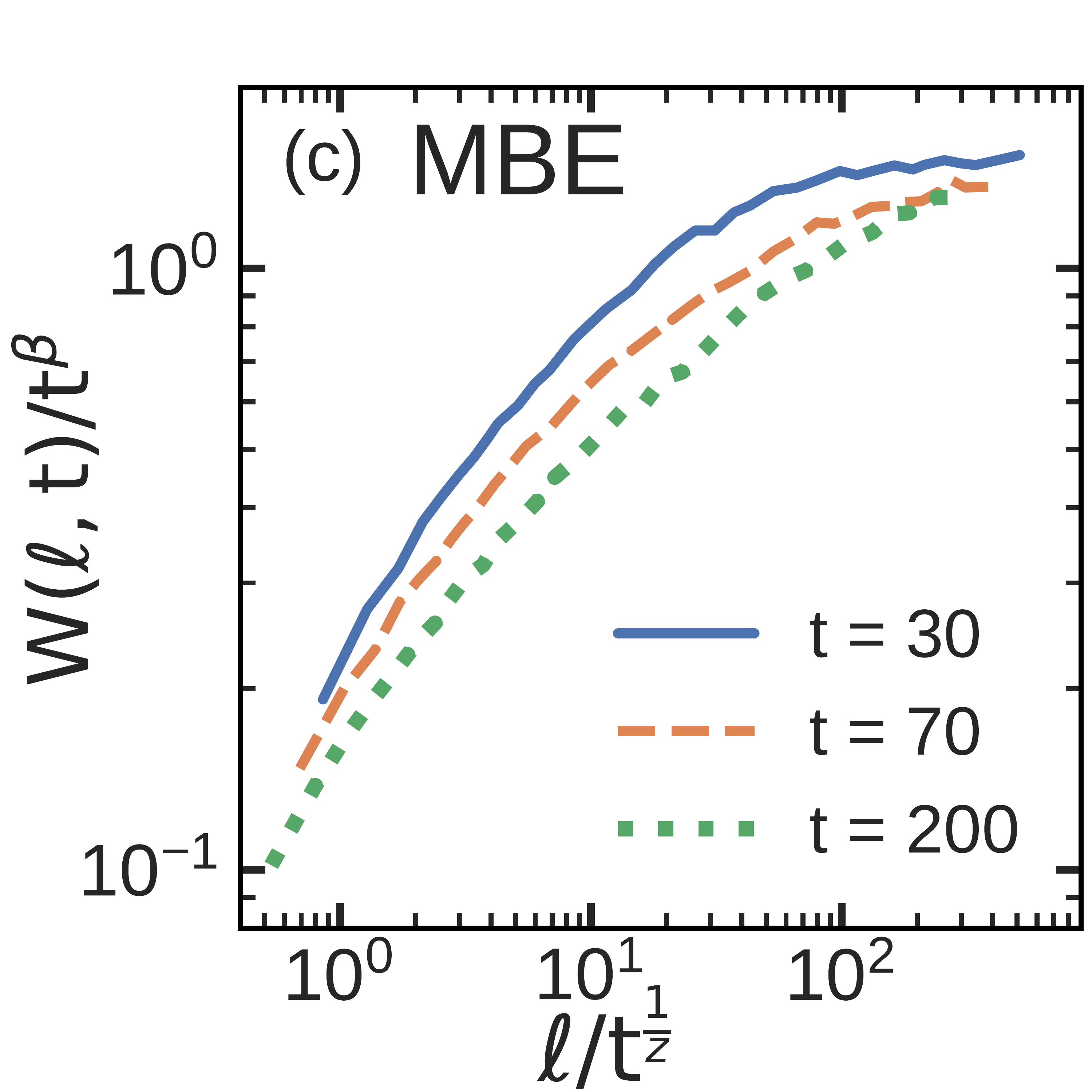} }
\resizebox{0.49\columnwidth}{!}{ \includegraphics{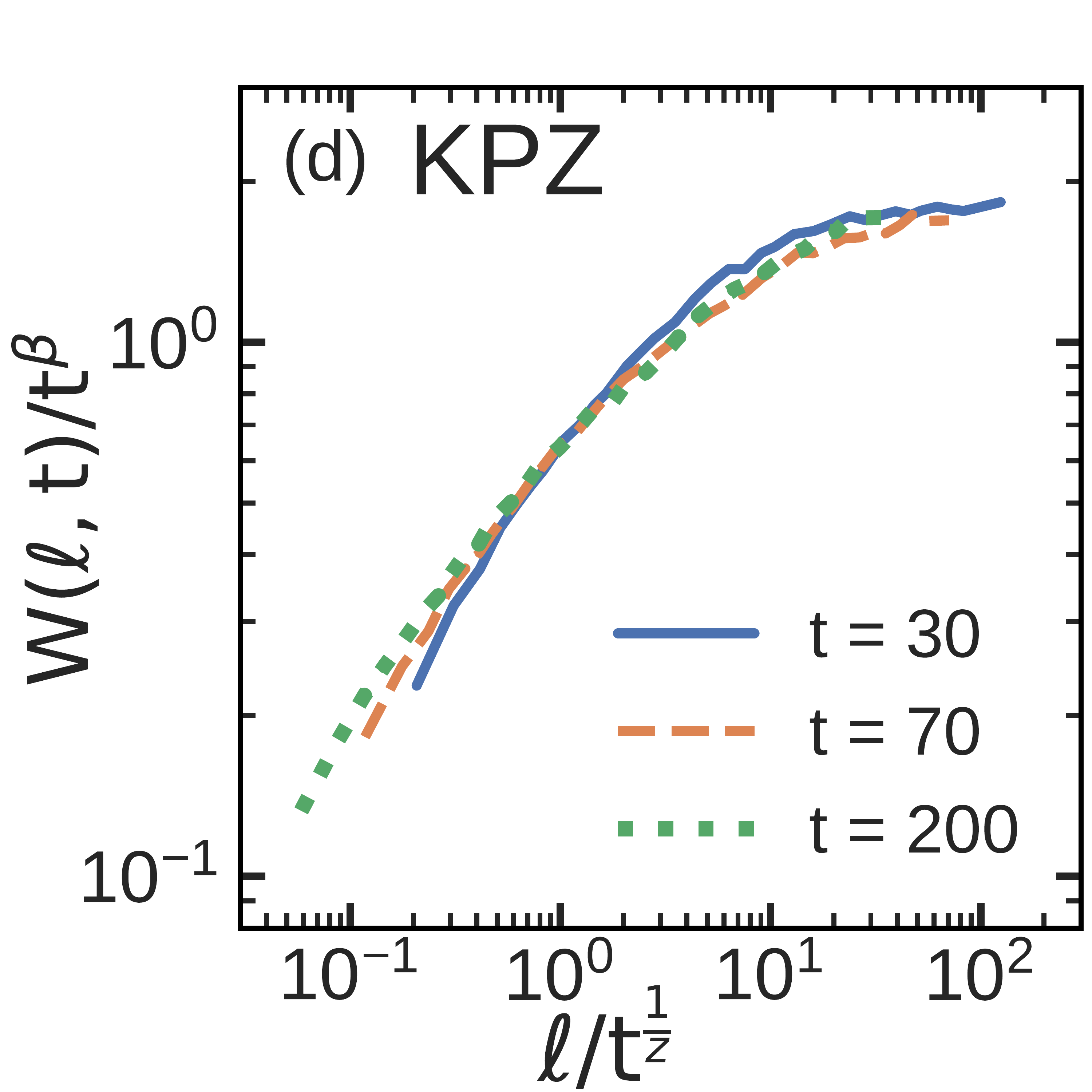} }
\caption{
Data collapse for line growth at high adhesion at three different times.
(a)  Using the structure factor (Eq.~\ref{eq:9}) and MBE exponents, $\alpha^\mathrm{MBE}  \! = \!  \frac{3}{2}$ and $z^\mathrm{MBE}  \! = \!  4$.  (b) With KPZ exponents, $\alpha^\mathrm{KPZ} \!  = \!  \frac{1}{2}$ and $z^\mathrm{KPZ} \! = \!  \frac{3}{2}$. The $y$-axis is scaled with  the factor $(\mathrm{k} \mathrm{t}^{\frac{1}{\mathrm{z}}})^{\frac{3}{2}}$ to have the same range as MBE scaling.
(c) Using the Family-Vicsek relation for width (Eq.~\ref{eq:1})  with MBE exponents, $\beta^\mathrm{MBE} \! = \! \frac{3}{8}$ and $z^\mathrm{MBE} \! = \!  4$,  and (d) with KPZ exponents, $\beta^\mathrm{KPZ} \! = \!  \frac{1}{3}$ and $z^\mathrm{KPZ} \! = \!  \frac{3}{2}$.  For units, see Table~S1.
}
\label{fig:fig104}
\end{figure}

We first consider the growth of linear fronts at two different adhesion strengths, $10$ (weak) and $2000$ (strong), see Table~S1 for parameters. The initial configurations had a line of 240 cells, and the final populations consisted of $\approx$200,000 cells. Snapshots are shown in Fig.~\ref{fig:fig101}. As this figure shows, increasing the cell-cell adhesion changes the morphology of the colony and the interface. Figure~\ref{fig:fig102}a shows that the  interfaces grow at constant velocities with the growth rate decreasing with increasing adhesion. 

Next, the box counting method was utilized to determine the fractal dimensions ($d_\mathrm{f}$) of the interfaces for both line configurations (triangles) and radially growing (circles) systems.  Figure~\ref{fig:fig102}b shows that $d_\mathrm{f}$ is in the same range for all simulations. In both geometries, however, $d_\mathrm{f}$ is slightly higher for the case of strong adhesion. This is consistent with the experiments of Torres Hoyos \textit{et al.}~\cite{hoyos2018geometrical} who reported smaller  $d_\mathrm{f}$ for malignant and invasive cancer cells as compared to the benign and more solid (higher adhesion) tumors.

Figures~\ref{fig:fig104}a-d show the scaling results for line growth both through the width function (Eq.~\ref{eq:1}) and the structure factor (Eq.~\ref{eq:9}). As the figure shows, both approaches
suggest KPZ dynamics; using the Family-Vicsek relation for the structure factor (Eq.~\ref{eq:9}) displays slightly better collapse for KPZ than for MBE, Figs.~\ref{fig:fig104}a,b,  and using width, Eq.~\ref{eq:1}, the data collapses to a single function using the KPZ exponents (Fig.~\ref{fig:fig104}d).

Figure~S1 shows that interface roughness, $w(t)$, follows a power-law $t^\beta$ with $\beta^\mathrm{weak} \! = \! 0.28 \pm 0.01$  and $\beta^\mathrm{strong}  \! = \! 0.25 \pm 0.02$ for weak and strong adhesion, respectively.
Figures~S2 and S3 show the local and global roughness exponents. As the figures show, determining the roughness exponent is  questionable especially in the case of weak adhesion. The value $\alpha_\mathrm{loc}^\mathrm{strong} \! = \!  0.62 \pm 0.02$,  was calculated using width, and  $\alpha_\mathrm{glob}^\mathrm{weak} \! = \!  0.75 \pm 0.04$ and $\alpha_\mathrm{glob}^\mathrm{strong} \! = \!  0.52 \pm 0.02$,  using the structure factor. As the results show, $\alpha_\mathrm{glob}$ decreases with increasing adhesion. 

\begin{figure}
\resizebox{0.49\columnwidth}{!}{ \includegraphics{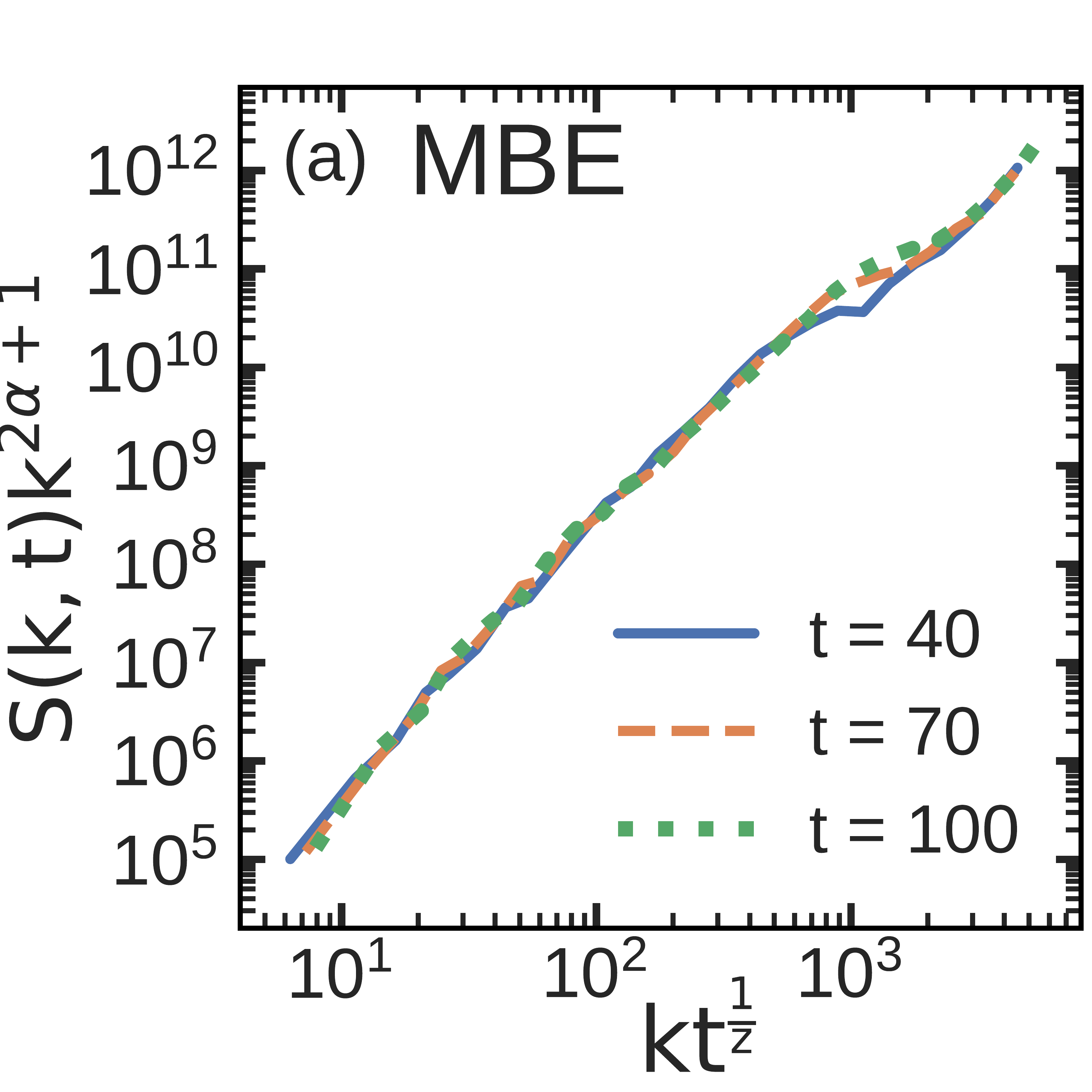} }
\resizebox{0.49\columnwidth}{!}{ \includegraphics{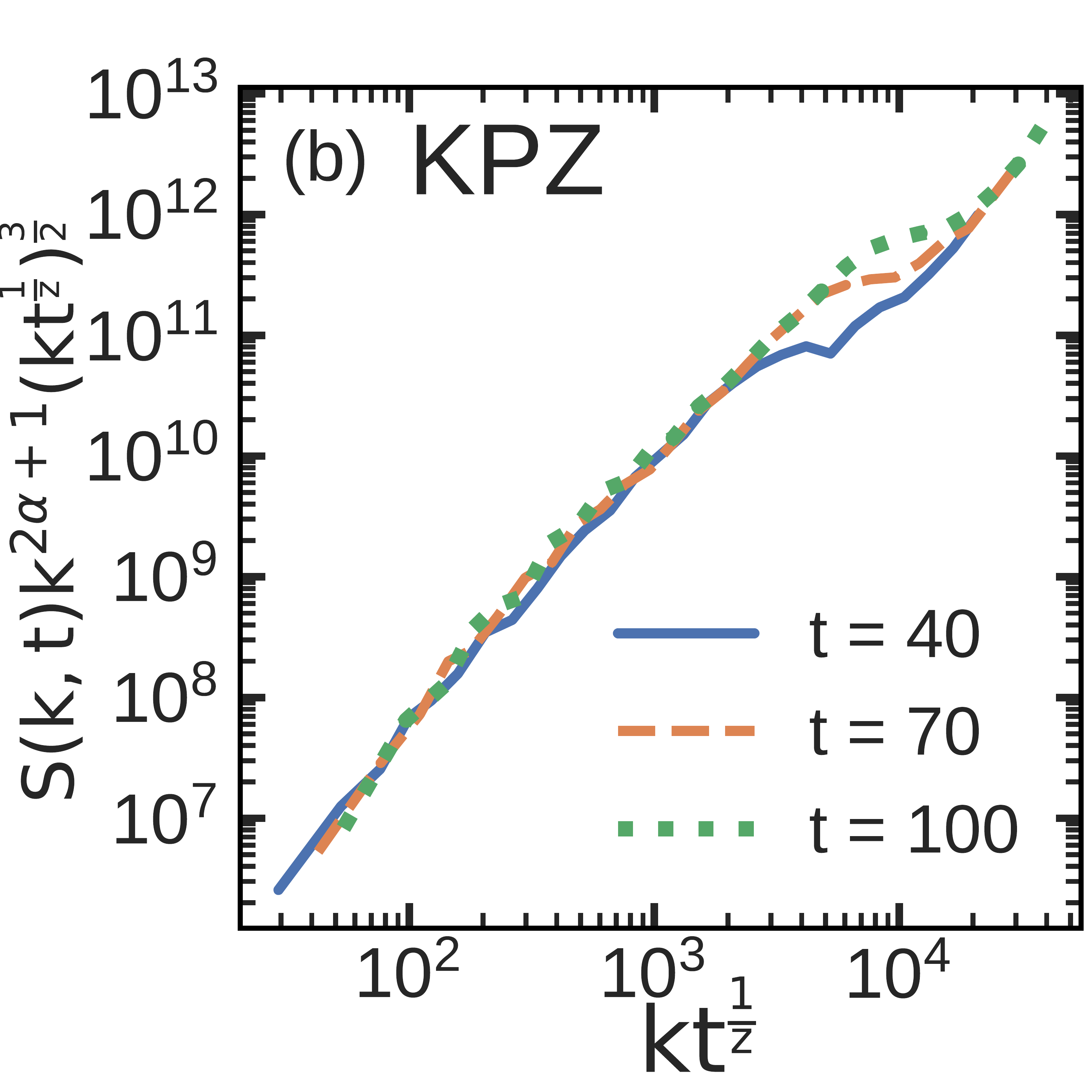} }
\resizebox{0.49\columnwidth}{!}{ \includegraphics{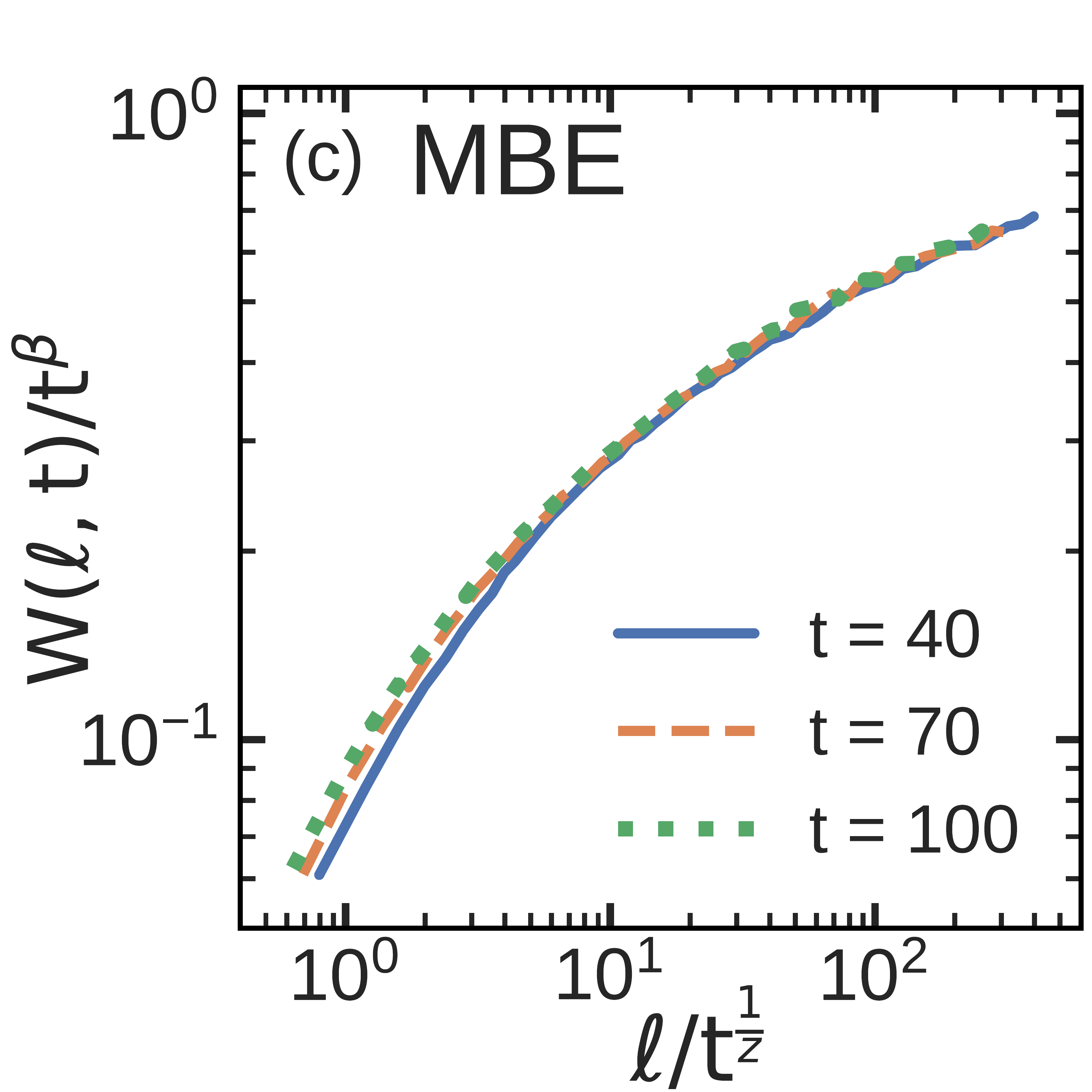} }
\resizebox{0.49\columnwidth}{!}{ \includegraphics{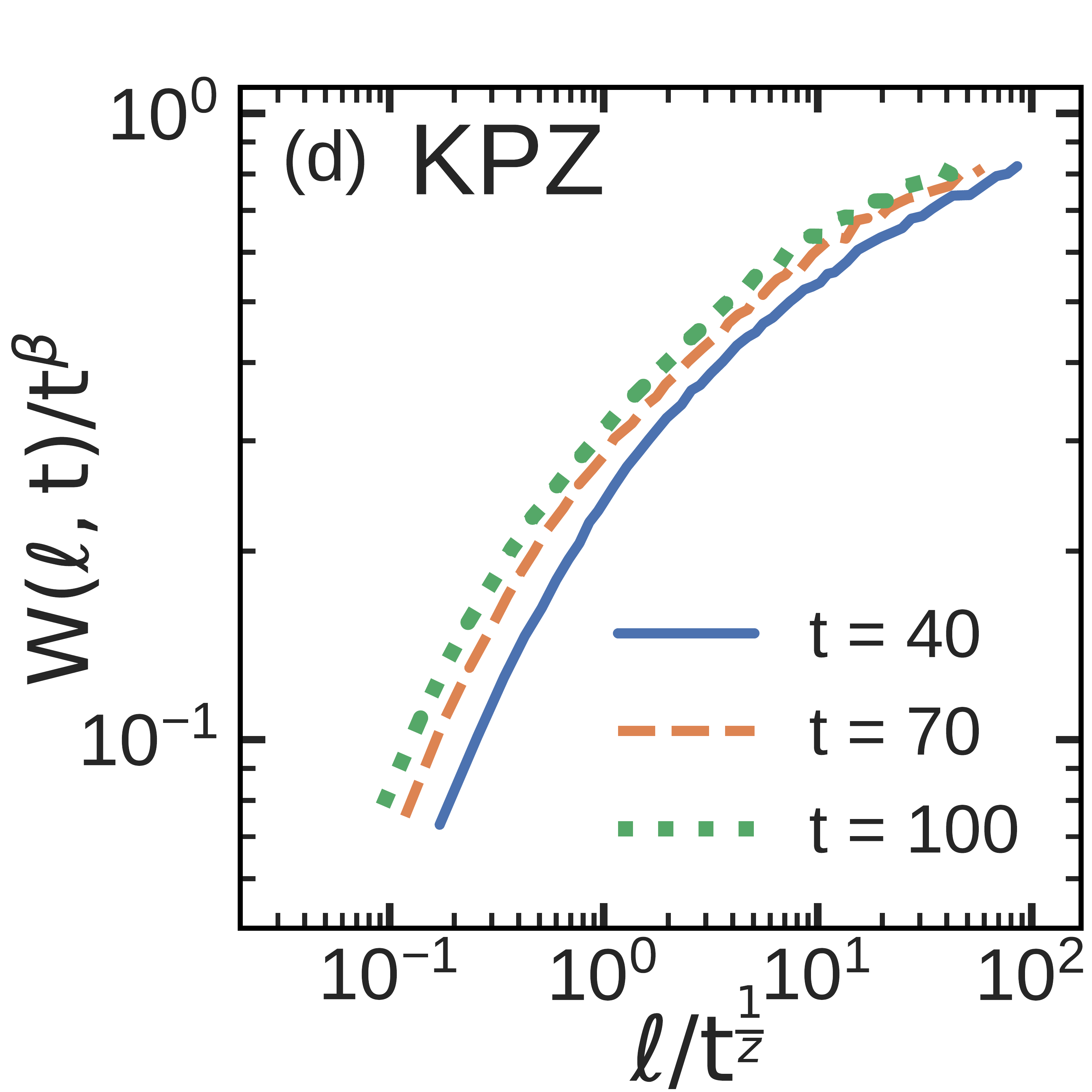} }
\caption{
Data collapse for the radially growing interface at low adhesion
at three different times. (a) Using the structure factor (Eq.~\ref{eq:9})
with MBE, $\alpha^\mathrm{MBE} \! = \! \frac{3}{2}$ and $z^\mathrm{MBE} \! = \! 4$ and (b) KPZ exponents, $\alpha^\mathrm{KPZ} \! = \! \frac{1}{2}$ and $z^\mathrm{KPZ} \! = \! \frac{3}{2}$.  The $y$-axis is 
scaled with the factor $(\mathrm{k} \mathrm{t}^{\frac{1}{\mathrm{z}}})^{\frac{3}{2}}$ to have the same range as MBE scaling. (c)  Using the Family-Vicsek relation for width (Eq.~\ref{eq:1}) with MBE exponents, $\beta^\mathrm{MBE} = \frac{3}{8}$ and $z^\mathrm{MBE} \! = \! 4$, and (d) with KPZ exponents, $\beta^\mathrm{KPZ} \! = \! \frac{1}{3}$ and $z^\mathrm{KPZ} \! = \! \frac{3}{2}$.  For units, see Table~S1.}
\label{fig:fig103}
\end{figure}

Next, we focus on radially expanding isotropic fronts. The initial configuration was one cell at the center of the simulation box and the final populations were $\approx$200,000 cells. We define $R_\mathrm{i}(t)$ to be the distance from the center of mass of the colony to the $i$th site at the interface. Snapshots at different times and adhesion strengths are shown in Fig.~\ref{fig:fig101}; increasing the adhesion between the cells changes the colony morphology and increases overhangs. The  radii grow at constant velocities, $ \langle v \rangle \! = \! 4.24 \pm 0.01 $ for weak and $ \langle v \rangle  \! = \! 2.15 \pm 0.01 $ for strong adhesion, Fig.~\ref{fig:fig102}a. As in the linear case, increasing the adhesion causes the front velocities to decrease.
Again, the fractal dimension (Fig.~\ref{fig:fig102}b)  shows slight increase with increasing adhesion,
$d_\mathrm{f}^\mathrm{weak} \! = \! 1.13 \pm 0.01$ and  $d_\mathrm{f}^\mathrm{strong} \! = \! 1.21 \pm 0.01$.

Figures~\ref{fig:fig103}a-d show the scaling results for radial growth both through the width function (Eq.~\ref{eq:1}) and structure factor (Eq.~\ref{eq:9}). Using the Family-Vicsek relation for width 
(Fig.~\ref{fig:fig103}c) with MBE exponents displays good collapse. In addition,  the structure factors at different times show good collapse with MBE exponents.

Figure~S4 shows that fitting the width vs time gives the growth exponents $\beta^\mathrm{weak} \! = \! 0.40 \pm 0.04$ and $\beta^\mathrm{strong} \! = \! 0.42 \pm 0.06$, For varying adhesion strengths, the local roughness exponent, $\alpha_\mathrm{loc}$, are all within the same range: $\alpha_\mathrm{loc}^\mathrm{weak} \! = \! 0.66 \pm 0.01$ and $\alpha_\mathrm{loc}^\mathrm{strong} \! = \! 0.7 \pm 0.01$, Fig.~S5. The scaling exponent $\alpha_\mathrm{glob}$, measured from  the structure factor, shows a decrease with increasing adhesion strength,  $\alpha_\mathrm{glob}^\mathrm{weak} \! = \! 0.95 \pm 0.04$ and $\alpha_\mathrm{glob}^\mathrm{strong} \! = \! 0.71 \pm 0.02$, Fig.~S6.

Finally, to test the generality of the above observations,  cell-medium friction, intermembrane friction, cell division rules, and the number of cell types with different stiffness were tested using the radially growing system. Figure~S7 shows that $\alpha_\mathrm{glob}$, is insensitive to changes in factors such as cell-medium friction, intermembrane friction, cell division rules, and the number of cell types with different stiffness. These results suggest that dynamics of the radially growing colonies are well described by MBE-like scaling when at weak cell-cell adhesion.

In conclusion, we have demonstrated that weak cell-cell adhesion and an isotropically growing colony display MBE-like scaling for the boundary roughness. This is consistent with the experiments of Br\'u \textit{et al.}~\cite{Bru2003}; we digitized their data and show its scaling in Fig.~S8. In contrast, a colony growing 
from a single line of cells, and with strong cell-cell adhesion displays KPZ-like scaling. This is in agreement with the experiments of Huergo \textit{et al.}~\cite{Huergo2010}. In all the studied cases the fractal dimensions of the interface are within the range 1.13--1.26, which is also consistent with prior experiments~\cite{Bru2003,Huergo2010,Huergo2011,Huergo2012,galeano2003dynamical}, regardless of the growth scaling behavior. 

All the studied cases show linear growth with constant velocity. This is consistent with experiments:
Br\'u \textit{et al.} reported  radially spreading cell colonies to exhibit exponential growth in the early stages
followed by linear growth~\cite{Bru2003}.  Huergo \textit{et al.} showed that once the number of cells exceeds 700–1,000, radially spreading  colonies grow with a constant velocity~\cite{Huergo2011,Huergo2012}. 
Constant velocity also applies for line growth~\cite{Huergo2010}.  Physically, these data suggest that cells are partially contact-inhibited and that most  activity occurs within a limited band along the interface challenging the notion of Gomperzian growth of cancer~\cite{Laird1964-xr}. Experiments by Costa \textit{et al.} indicate that \textit{in vitro} cultivated cells may exhibit sigmoidal growth~\cite{costa2015universal}.

Finally, it can be assumed that the first multi-cellular life-forms were rather simple cell colonies. 
It is interesting to speculate that if cell-colony growth dynamics would be rigidly confined to a single universality class, adaptive evolution would likely be significantly harder in contrast to more versatile growth dynamics.  This is consistent with earlier results in the sense that very rich growth behaviour and diverse tissues appear with modest changes, and, e.g., non-trivial dependencies in initial conditions, nutrients, apoptosis, disorder and mechanical forces from various sources cause changes in both quantitative and qualitative behaviors, see, e.g., Refs.~\cite{Lecuit2007-fk,Tambe2011,azimzade2019effect,Madhikar2020,Madhikar2021-ko}.

\begin{acknowledgments}
 We thank the Natural Sciences and Engineering Research Council of Canada (MK), Canada Research Chairs Programs (MK) and Western University's Science International Engagement Fund (MM) for financial support. The use of the computational resources provided by the Finnish Grid and Cloud Infrastructure FGCI, funded by the Academy of Finland, is gratefully acknowledged.

\end{acknowledgments}


%

\end{document}


\title{Supplemental Material: \\Epithelial Tissue Growth Dynamics: Universal or Not?}

\author{Mahmood Mazarei}
\affiliation{Department of Mathematics,
  The University of Western Ontario, 1151 Richmond Street, London,
  Ontario, Canada  N6A\,5B7}
\affiliation{The Centre for Advanced Materials and Biomaterials Research,
  The University of Western Ontario, 1151 Richmond Street,
  London, Ontario, Canada N6A\,3K7}

\author{Jan {\AA}str{\"o}m}
\affiliation{CSC Scientific Computing Ltd, K{\"a}gelstranden 14, 02150
  Esbo, Finland}

\author{Jan Westerholm}
\affiliation{Faculty of Science and Engineering, {\AA}bo Akademi
University, Vattenborgsv\"agen 3, FI-20500, {\AA}bo, Finland
}

\author{Mikko Karttunen}
\affiliation{The Centre for Advanced Materials and Biomaterials Research,
  The University of Western Ontario, 1151 Richmond Street,
London, Ontario, Canada N6A\,3K7}
\affiliation{Department of Chemistry,  The University of Western Ontario, 1151 Richmond Street, London,
  Ontario, Canada N6A\,5B7
}
\affiliation{Department of Physics and Astronomy,
  The University of Western Ontario, 1151 Richmond Street, London,
  Ontario, Canada  N6A\,3K7}

\date{\today}

\maketitle

\section{The \textit{CellSim3D} model}

\textit{CellSim3D}~\cite{Madhikar2018,Madhikar2020} is a 3D generalization of the 2D model originally introduced by Mkrtchyan, {\AA}str\"om, and Karttunen~\cite{Mkrtchyan2014}. 
In \textit{CellSim3D}
cells are defined as a set of interconnected nodes on the surface of a spherical C180 fullerene. The force field consists of bonded and non-bonded forces defined as
\begin{equation}
\label{eq:12}
\mathbf{F} = m \mathbf{ \ddot{r} } = \mathbf{F}^\mathrm{B} + \mathbf{F}^\mathrm{\theta} + \mathbf{F}^\mathrm{P} + \mathbf{F}^\mathrm{R} + \mathbf{F}^\mathrm{A} + \mathbf{F}^\mathrm{F} + \eta, 
\end{equation} 
where $\mathbf{F}^\mathrm{B}$ is a damped harmonic oscillator force between neighboring nodes, $\mathbf{F}^\mathrm{\theta}$ is the angle force which preserves cell curvature. These two forces are bonded, and the rest are non-bonded. $\mathbf{F}^\mathrm{R}$ and $\mathbf{F}^\mathrm{A}$ are the repulsive and attractive force between C180 nodes in different cells, respectively. They are approximated by harmonic potentials with different spring constants and cutoffs. The model has been shown to produce versatile behavior in agreement with experiments~\cite{Mkrtchyan2014,Madhikar2018,Madhikar2020,Madhikar2021-ko}. We discuss some of the main features below.

Adhesion forces in living cells are generated by adhesive bonds between Cell Adhesion Molecules (CAMs) on the surface of neighbouring cells~\cite{vanRoy2008,Stemmler2008,Buckley1998}. Each node on the boundary of the simulated cells behaves as an adhesive site, approximating the average CAM behavior~\cite{Madhikar2018}. $\mathbf{F}^\mathrm{F}$ is the friction term decomposed into viscous drag due to the medium approximating the interactions between the cell and the extracellular matrix, and intermembrane friction, which  is proportional to the component of the relative velocity tangential to the surface of
the cells~\cite{Madhikar2018}. Together attractive, repulsive, and friction terms approximate inter-membrane interactions mediated by cell adhesion molecules~\cite{Murray1999,Edelman1983,Edelman1991,Schluter2014,Helvert2017}. Finally, $\mathbf{F}^\mathrm{P}$ is the growth force coming from the osmotic pressure within cells and the cell's internal pressure~\cite{Murray1999}, and $\eta \equiv \eta(\mathbf{x},t)$, is Gaussian white noise.  

In \textit{CellSim3D}, growth is induced by increasing the internal pressure, as determined by experimental measurements~\cite{Stewart2011}.  At each time step, the internal pressure is incremented, resulting in a increase in the pressure force~\cite{Mkrtchyan2013,Madhikar2018}. As a result, the cell volume gradually increases, $V = V_0 + \Delta V$, and once it exceeds the threshold value, $V_\mathrm{div}$, the cell can divide into two new symmetric or asymmetric daughter cells depending on the chosen division rule~\cite{morrison2006asymmetric}. In this work, we consider that the daughter cells are symmetric~\cite{kimble2005germline,imoto2011cell}. 

The cell division plane's position can be either through the cell's centre of mass or through another randomly chosen point inside the cell. The random point might be different for each cell, resulting in asymmetric division.  The division plane's orientation can also be set using a variety of rules. According to Hertwig's rule, the division plane is perpendicular to the longest axis of an ellipsoidal cell and passes through the centre of mass~\cite{Bosveld2016}. Errera's rule asserts that the division plane must contain the shortest path travelling through the centre of mass of a cell~\cite{Sahlin2010}. The division plane can also be randomly oriented. In \textit{CellSim3D}, in the case of full 3D tissue, the division plane vector is chosen randomly from a unit sphere. In the case of 2D epithelial tissue, the division plane is sampled from a unit circle in the epithelial plane~\cite{Madhikar2020}. The parameters used in the simulations are listed in Table~\ref{tab:table1}. These parameters are based on the properties of HeLa cells for parameter mapping. For further details and their derivation about the \textit{CellSim3D} force field, methodology, parameter mapping, and code implementation, please see Refs.~\cite{Mkrtchyan2014,Madhikar2018,Jorgensen2004}.
\section{Analysis methods}
\begin{figure}[h!]
\resizebox{0.4\columnwidth}{!}{ \includegraphics{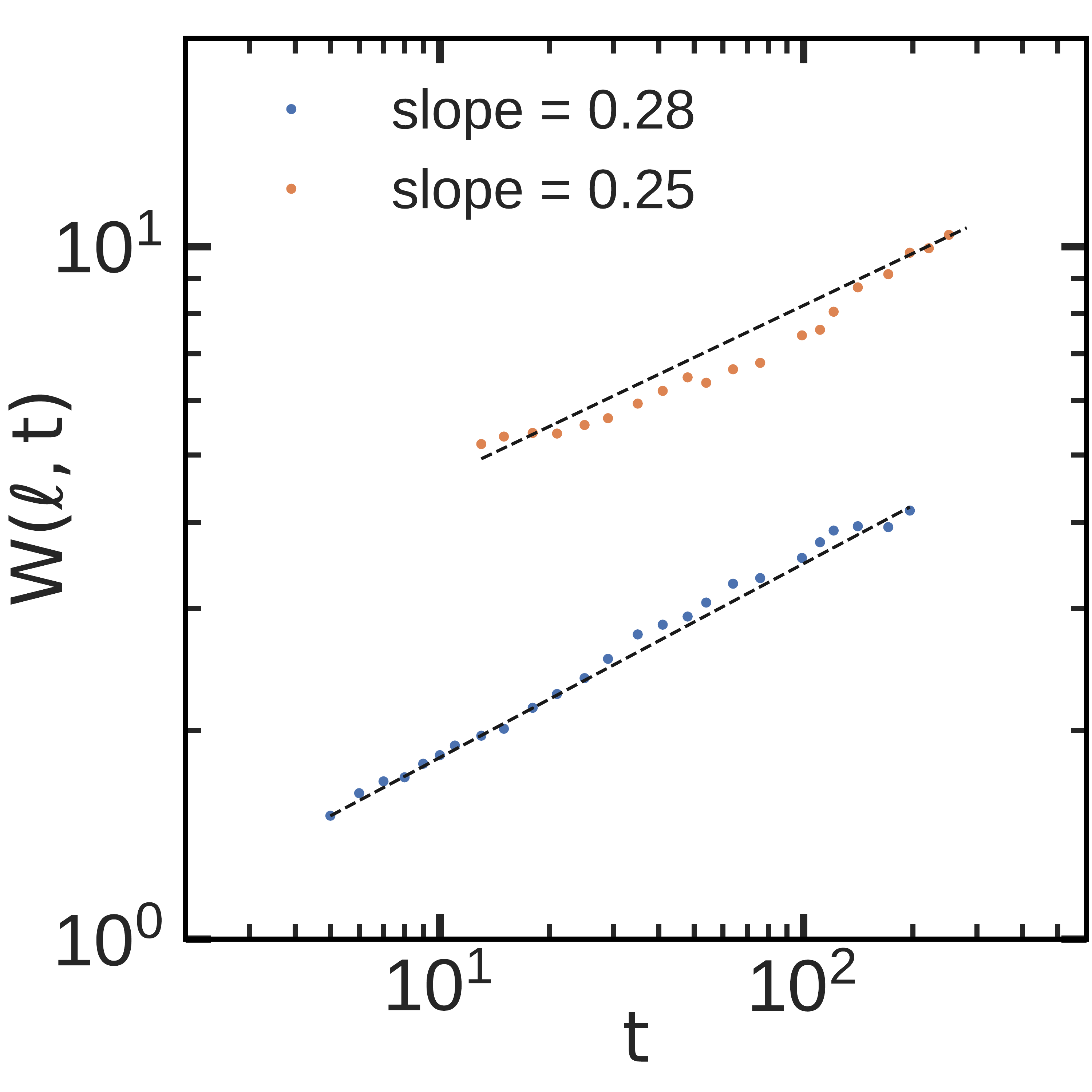} }
\caption{ Log-log plot of the colony interface width vs growth time of the linear interface for different adhesion strength, $10$ (blue) and $2,000$ (orange). Growth exponents are $\beta^\mathrm{weak} = 0.28 \pm 0.01$ and $\beta^\mathrm{strong} = 0.25 \pm 0.02$, respectively. For units, see Table~\ref{tab:table1}.
}
\label{fig:fig1}
\end{figure}

\begin{figure}[h!]
        \includegraphics[height=2.5in]{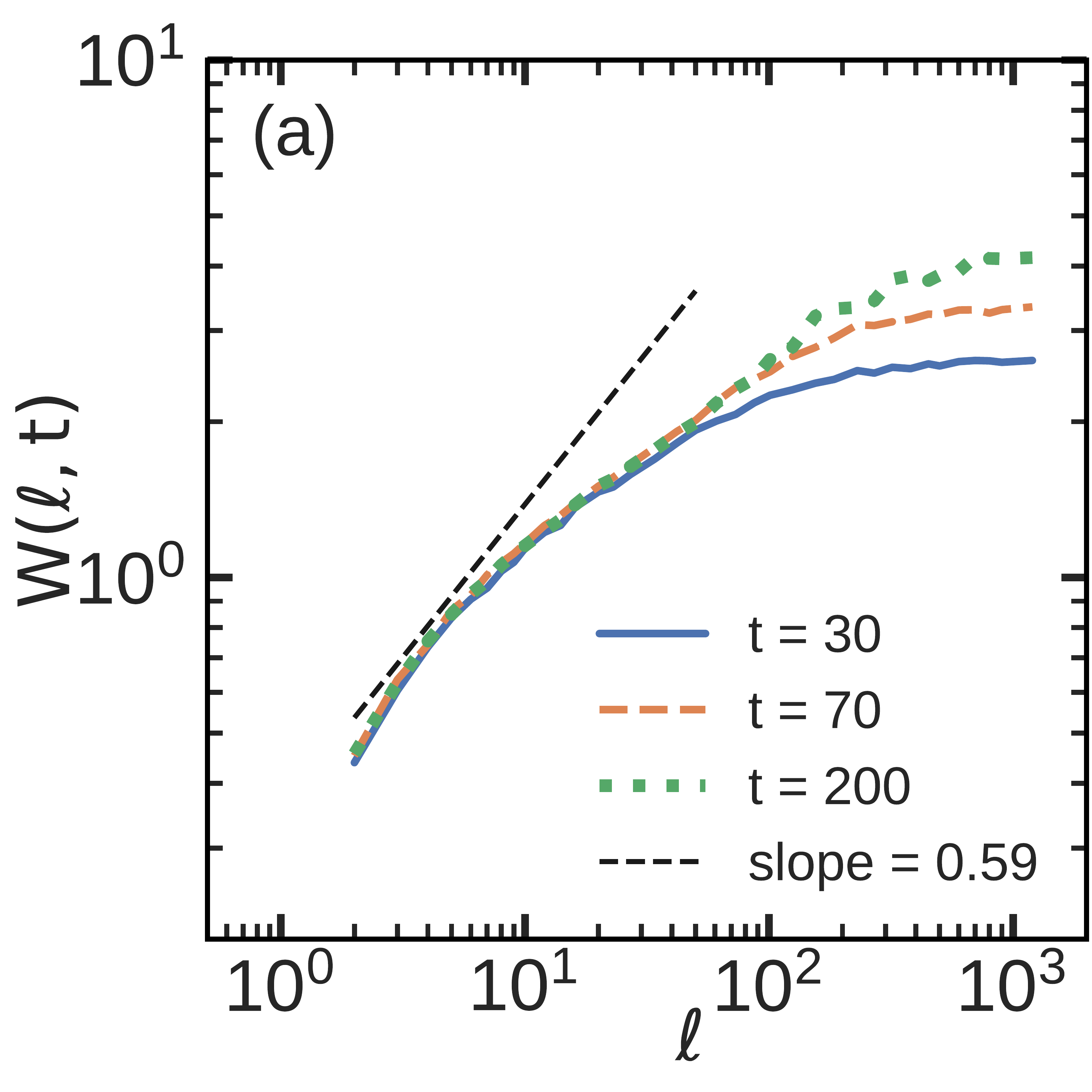}
        \includegraphics[height=2.5in]{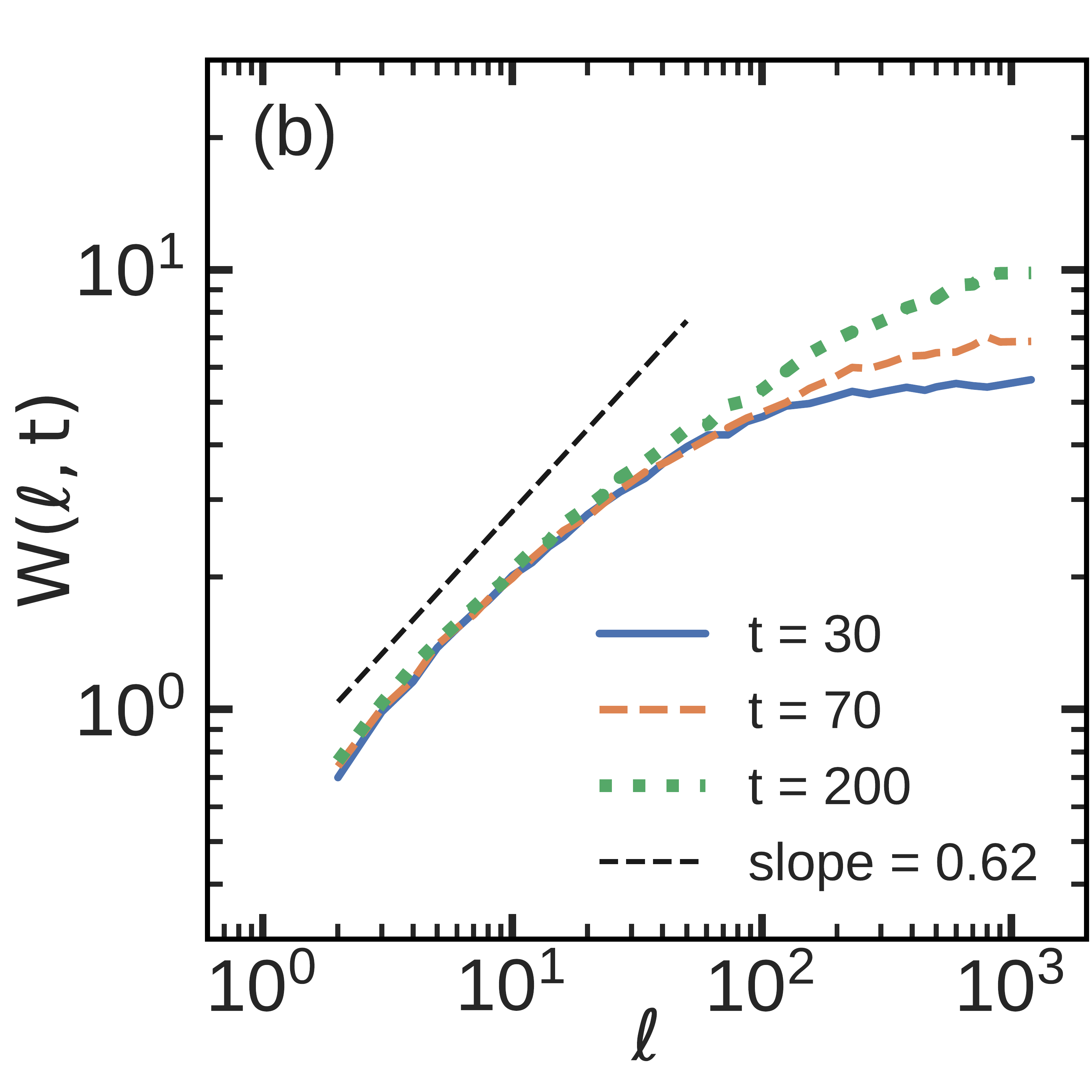}
        \caption{Log-log plot of the colony interface width of the linear interface vs length $l$ at three different times. (a) weak adhesion strength ($10$) with $\alpha_\mathrm{loc}^\mathrm{weak} = 0.59 \pm 0.01$ (b) strong adhesion strength ($2,000$) with $\alpha_\mathrm{loc}^\mathrm{strong} = 0.62 \pm 0.02$. For units, see Table~\ref{tab:table1}.
}
\label{fig:fig2}
\end{figure}

\begin{figure}[h!]
        \includegraphics[height=2.5in]{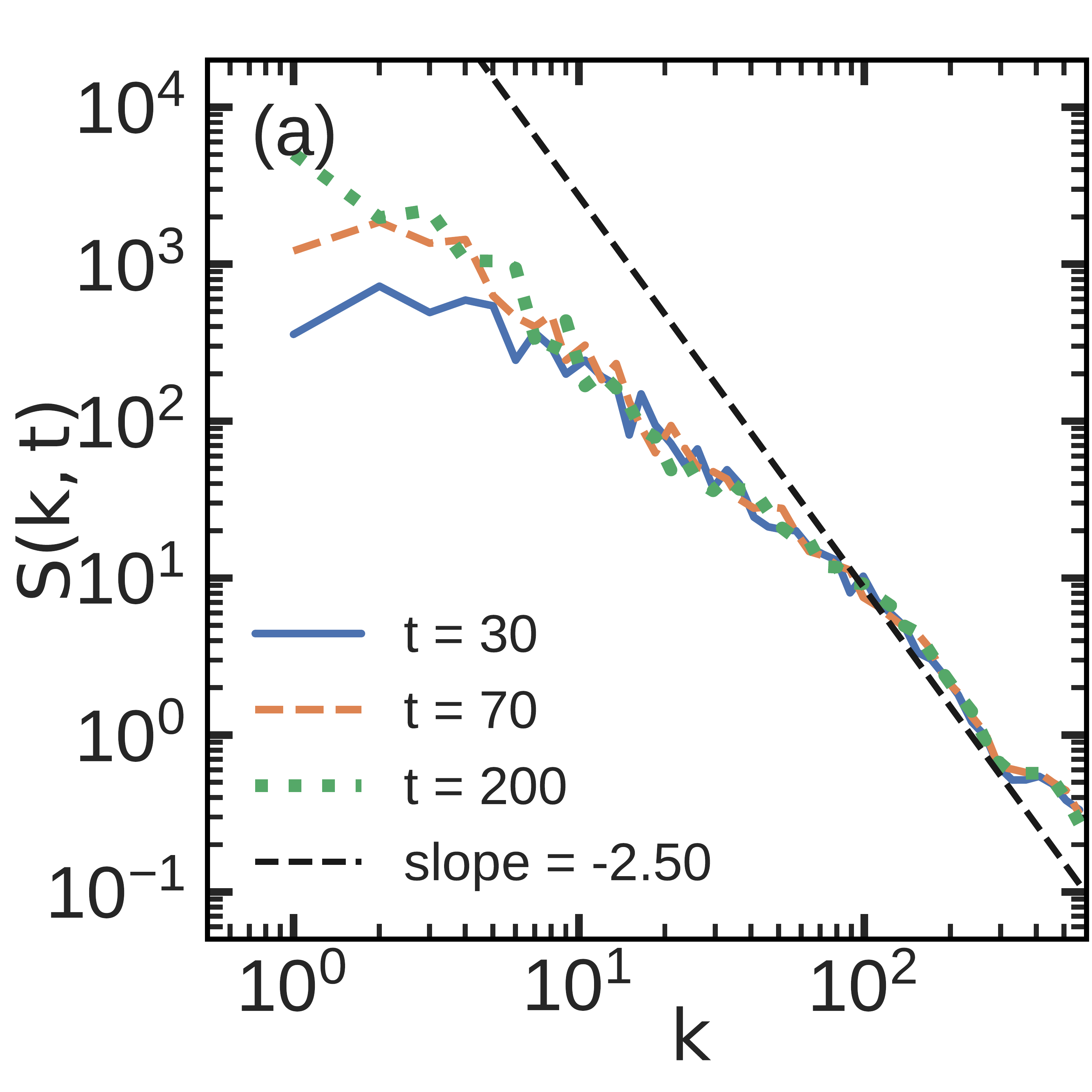}
        \includegraphics[height=2.5in]{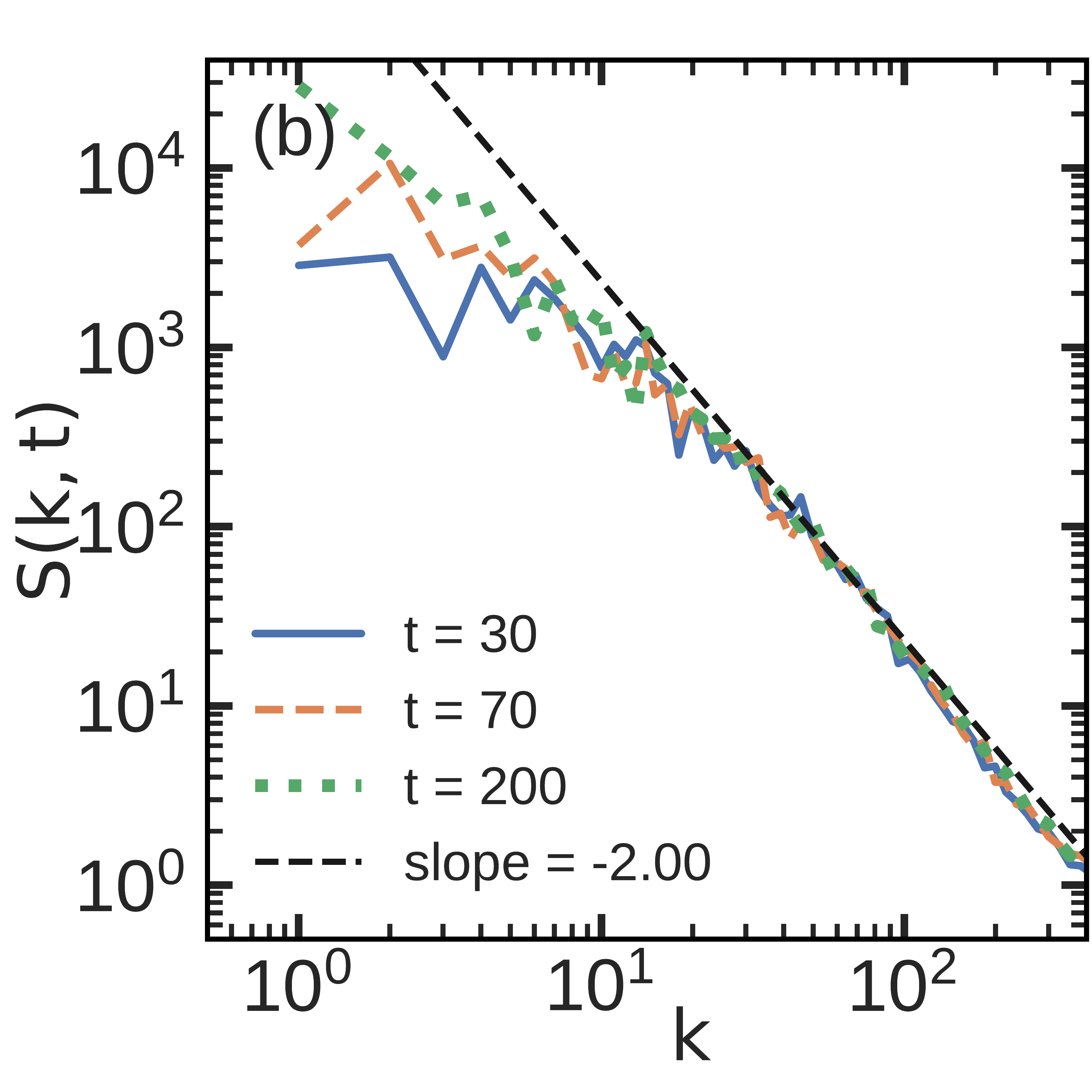}
\caption{ Log-Log plot of the structure factor at three different times for linear expanding interface. (a) weak adhesion strength ($10$) with $\alpha_\mathrm{glob}^\mathrm{weak} = 0.75 \pm 0.04$. (b) strong adhesion strength ($2,000$) with $\alpha_\mathrm{glob}^\mathrm{strong} = 0.52 \pm 0.02$. For units, see Table~\ref{tab:table1}.
}
\label{fig:fig3}
\end{figure}

\begin{figure}[h!]
\resizebox{0.4\columnwidth}{!}{ \includegraphics{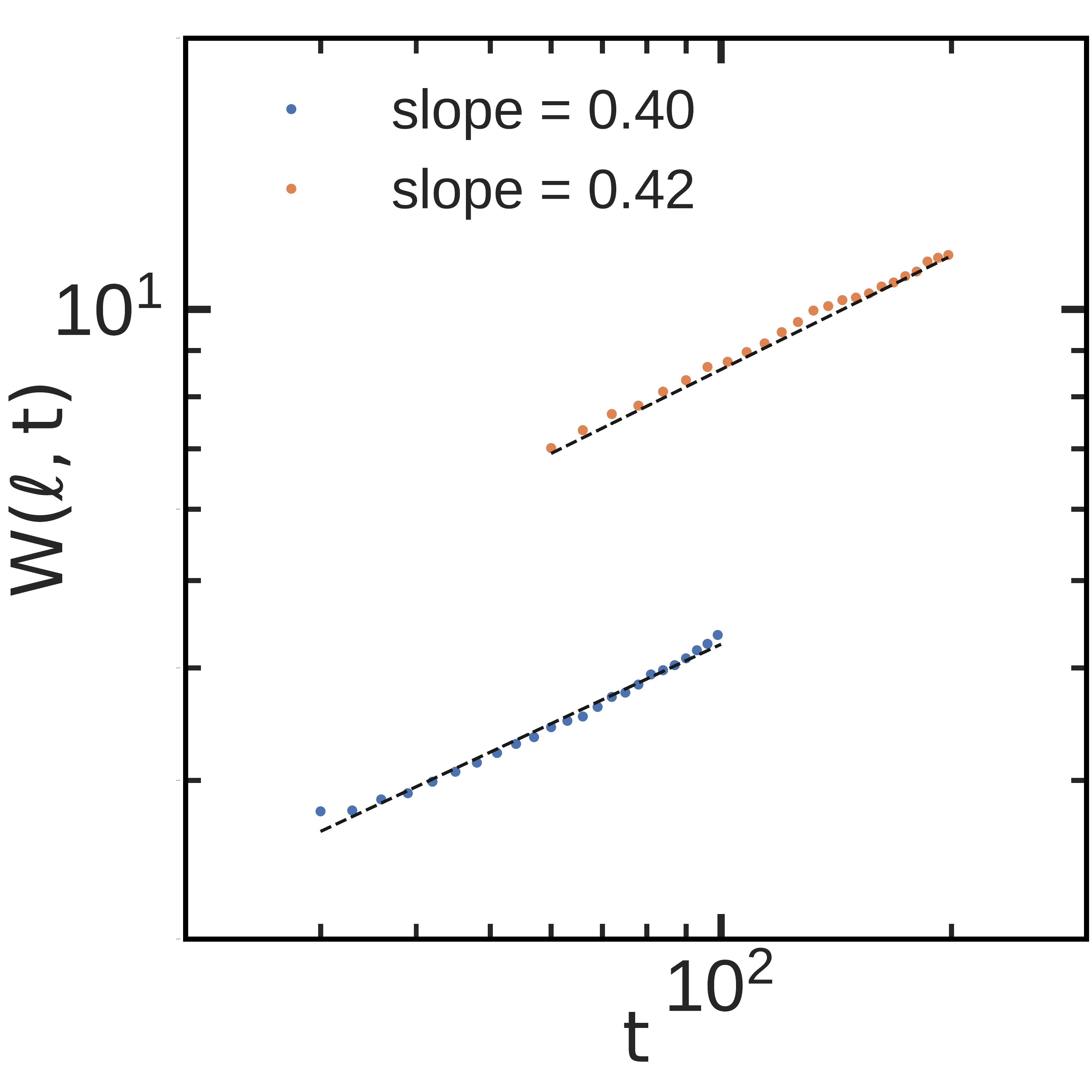} }
\caption{ Log-log plot of the colony interface width vs growth time of the radially expanding interface for different adhesion strength, $10$ (Blue) and $2,000$ (Orange). Growth exponents are $\beta^\mathrm{weak} = 0.40 \pm 0.04$ and $\beta^\mathrm{strong} = 0.42 \pm 0.06$, respectively.
}
\label{fig:fig4}
\end{figure}

\begin{figure}[h!]
        \includegraphics[height=2.5in]{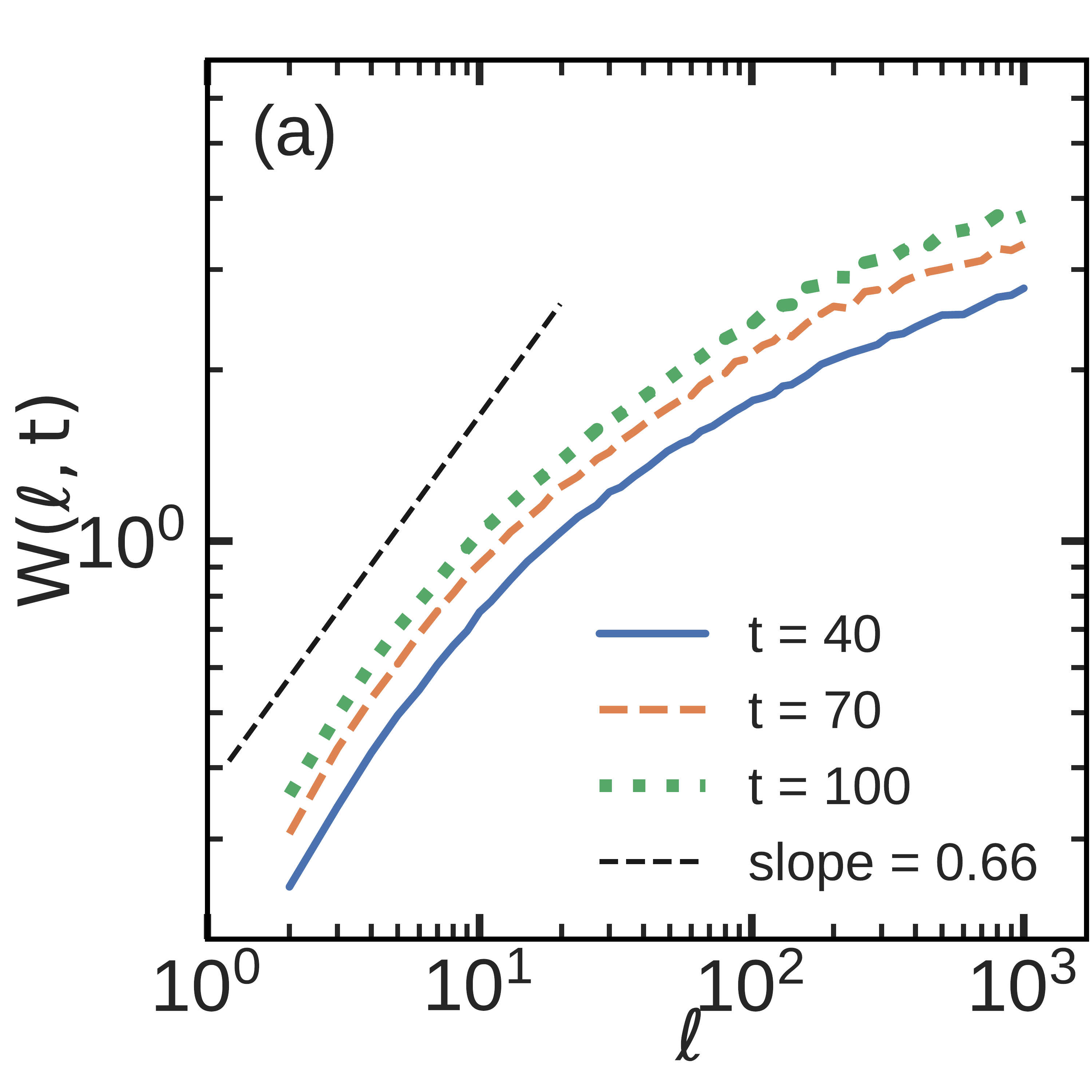}
        \includegraphics[height=2.5in]{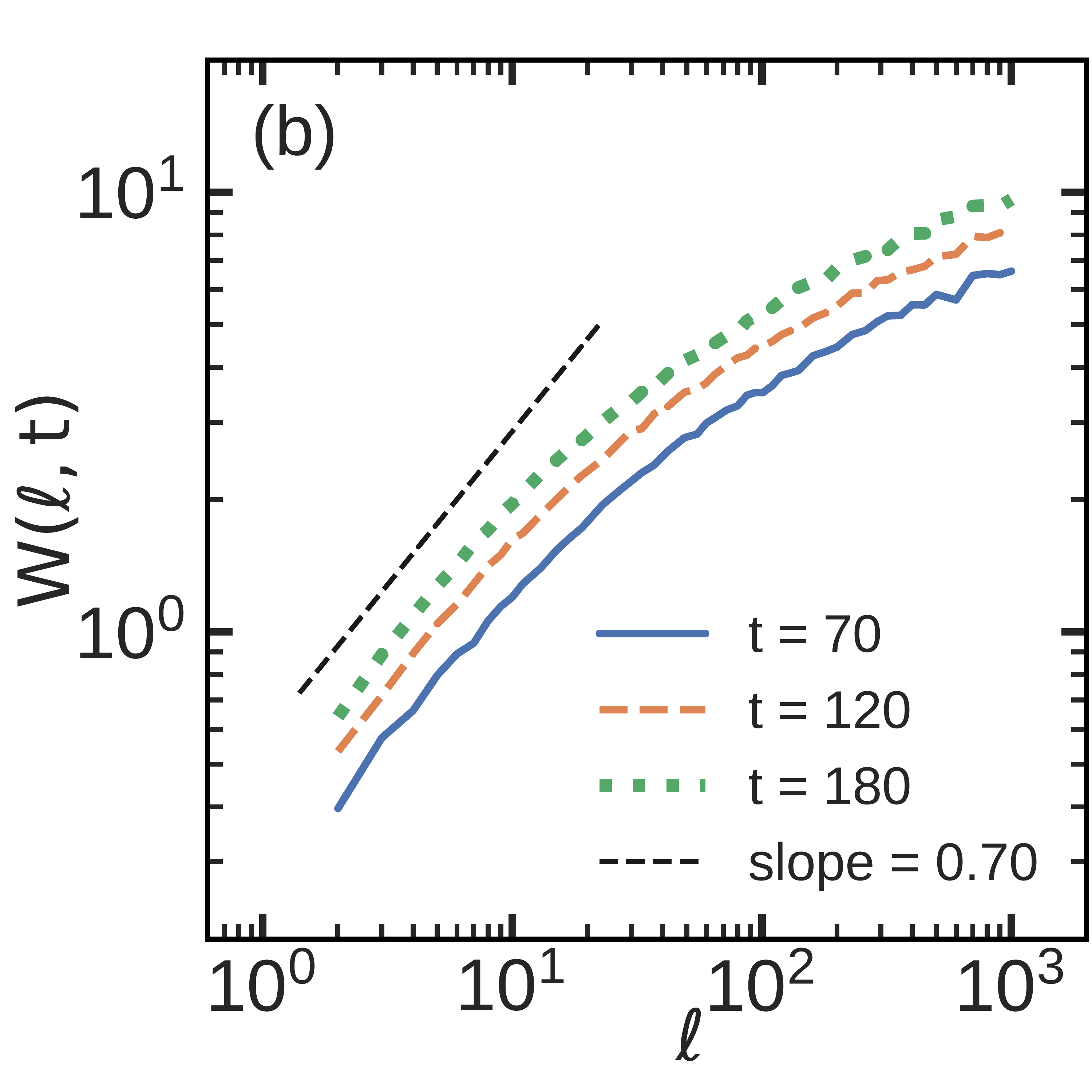}
\caption{Log-log plot of the colony interface width of the radially expanding interface vs length $l$ at different times. (a) Weak adhesion strength ($10$) with slope $\alpha_\mathrm{loc}^\mathrm{weak} = 0.66 \pm 0.01$, and (b) strong adhesion strength ($2,000$) with slope $\alpha_\mathrm{loc}^\mathrm{strong} = 0.70 \pm 0.01$. For units, see Table~\ref{tab:table1}.
}
\label{fig:fig5}
\end{figure}

\begin{figure}[h!]
        \includegraphics[height=2.5in]{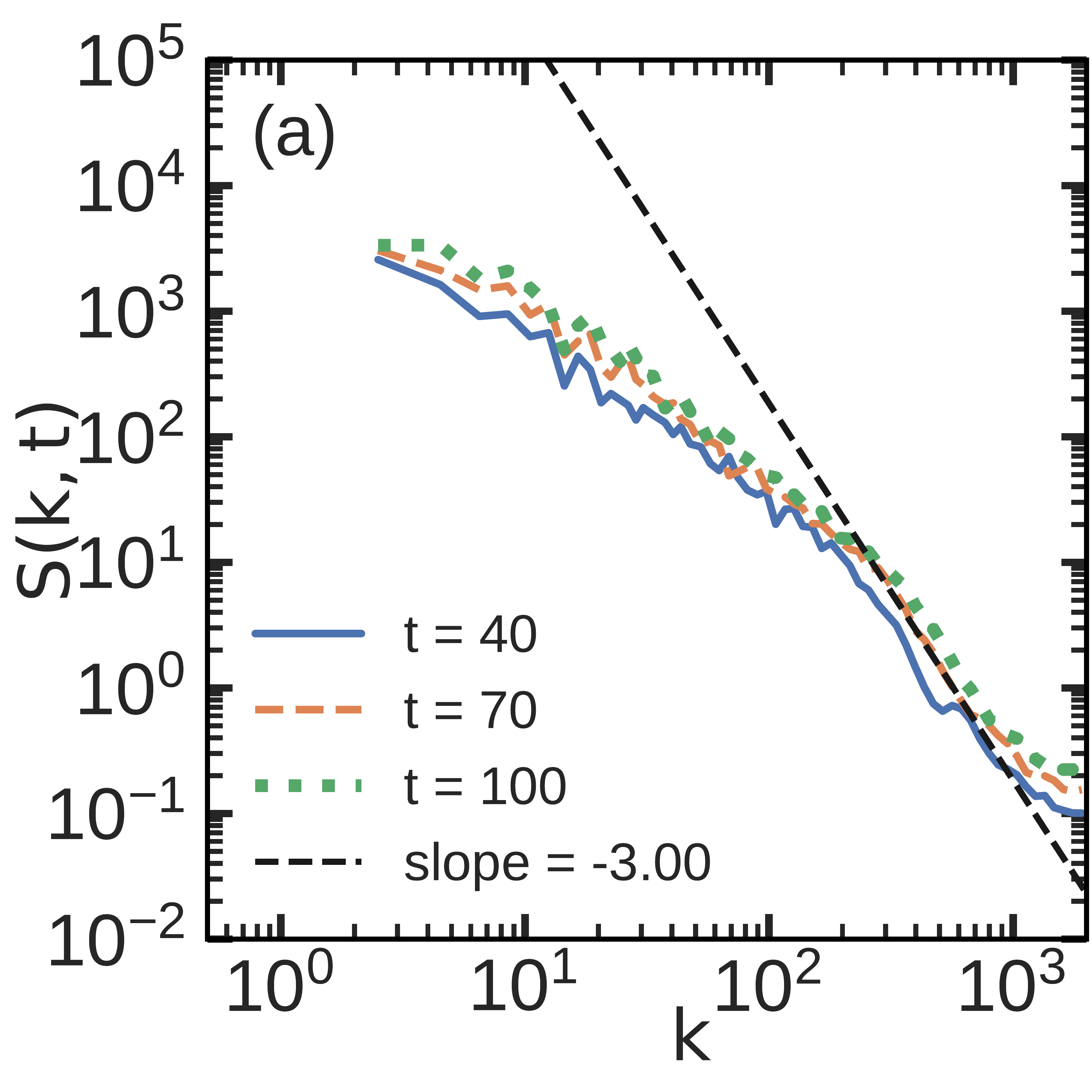}
        \includegraphics[height=2.5in]{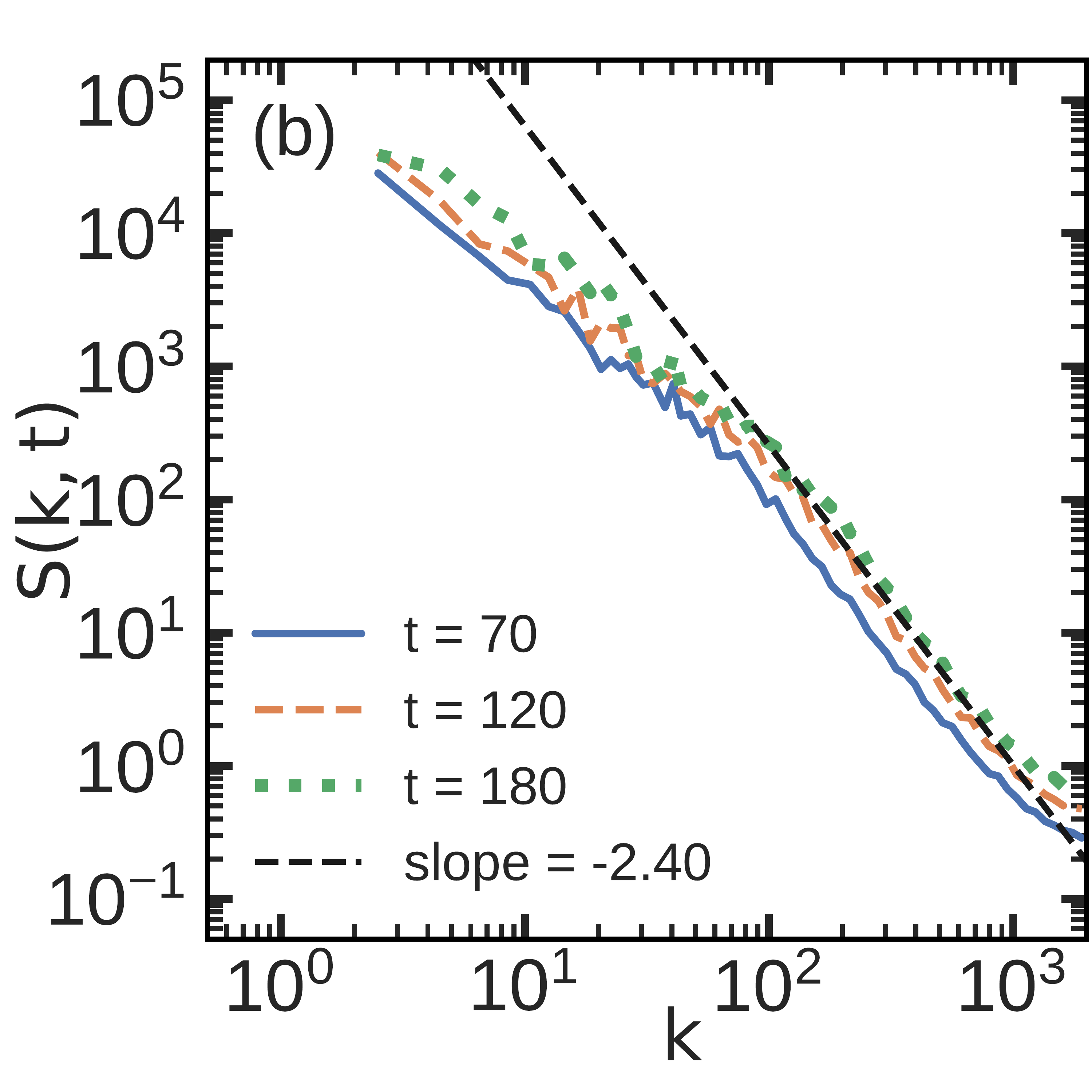}
\caption{Log-Log plot of the structure factor at three different times for the radially expanding interface. (a) weak adhesion
strength ($10$) with $\alpha_\mathrm{glob}^\mathrm{weak} = 0.95 \pm 0.04$, and (b) strong adhesion strength ($2000$) with $\alpha_\mathrm{glob}^\mathrm{strong} = 0.71 \pm 0.02$.
}
\label{fig:fig6}
\end{figure}

The interface local width function, $w(l,t)$, is defined as the standard deviation of the front height, $h(x,t)$, over a length scale $l$ at time $t$ as~\cite{Barabasi1995}
\begin{equation}
\label{eq:5}
w(l,t) = \bigg \lbrace \frac{1}{N} \sum_{i=1}^{N} [h_{i}(t) - \langle h_{i} \rangle_{l}]^{2} \bigg \rbrace _\mathrm{L}^{\frac{1}{2}},
\end{equation} 
For small length scales the width function behaves as follows 
\begin{equation}
\label{eq:1}
w(l,t) \sim l^\mathrm{\alpha_\mathrm{loc}}
\end{equation} 
where $\alpha_\mathrm{loc}$ is the local roughness exponent.
By replacing $l$ with the system size $L$, the value of $w(L,t)$ for $t\ll t_{s}$ , where $t_{s}$ is the roughness saturation time, is expected to increase as
\begin{equation}
\label{eq:2}
w(L,t) \sim t^{\beta}
\end{equation} 
where $\beta$ is the growth exponent~\cite{Barabasi1995}.
We supplement the universality class analysis by an examination of the structure factor, $S(k,t)$, of the interface,
\begin{equation}
\label{eq:3}
S(k,t) = \langle \hat{h}(k,t) \hat{h}(-k,t) \rangle ,
\end{equation}
The linear interface roughness, $w(t)$, follows the power-law $t^\mathrm{\beta}$ with $\beta^\mathrm{weak} \! = \! 0.28 \pm 0.01$, for weak adhesion and $\beta^\mathrm{strong} \! = \! 0.25 \pm 0.02$, for strong adhesion, Fig.~\ref{fig:fig1}. The local roughness exponents, $\alpha_\mathrm{loc}$, are within the same range for different adhesion strengths in the case of linear interface: $\alpha_\mathrm{loc}^\mathrm{weak} \! = \! 0.59 \pm 0.01$ for weak adhesion and $\alpha_\mathrm{loc}^\mathrm{strong} = \! 0.62 \pm \! 0.02$ for strong adhesion, Fig.~\ref{fig:fig2}, The global roughness exponent, $\alpha_\mathrm{glob}$, of the linear growing interface was calculated via structure factor analysis, $\alpha_\mathrm{glob}^\mathrm{weak} \! = \!  0.75 \pm 0.04$ and $\alpha_\mathrm{glob}^\mathrm{strong} \! = \!  0.52 \pm 0.02$, Fig.~\ref{fig:fig3}.
The width functions of the circular interface versus time $t^\mathrm{\beta}$ are displayed in Fig.~\ref{fig:fig4}: $\beta^\mathrm{weak} \! = \! 0.40 \pm 0.04$ for weak adhesion and $\beta^\mathrm{strong} \! = \! 0.42 \pm 0.06$ for strong adhesion.
The local roughness exponents, $\alpha_\mathrm{loc}$, for circular interface are $\alpha_\mathrm{loc}^\mathrm{weak} \! = \! 0.66 \pm 0.01$ for weak adhesion and $\alpha_\mathrm{loc}^\mathrm{strong} \! = \! 0.7 \pm 0.01$ for strong adhesion, Fig.~\ref{fig:fig5}. Finally, For circular interfaces, we find $\alpha_\mathrm{glob}^\mathrm{weak} \! = \! 0.95 \pm 0.04$ and $\alpha_\mathrm{glob}^\mathrm{strong} \! = \! 0.71 \pm 0.02$, Fig.~\ref{fig:fig6}.

\section{Analyses on Br\'u's data set}

Data from Fig.~3 of Br\'u \textit{et al.}~\cite{Bru2003} was digitized with the WebPlotDigitizer program~\cite{Rohatgi2020}. Figure~\ref{fig:fig9} shows the result from using the Family-Vicsek relation to scale the width function by MBE (Fig.~\ref{fig:fig9}a) and KPZ (Fig.~\ref{fig:fig9}b) exponents. MBE exponents show a better collaps than KPZ exponents, as Br\'u \textit{et al.} stated in their study. However, we cannot obtain a full collapse for the width functions at different times with MBE exponents, implying that the development dynamics of cell-colony interfaces do not perfectly fit into the MBE universality class. The similar conclusion can be made about Huergo \textit{et al.} results~\cite{Huergo2010,Huergo2011,Huergo2012}. Although they concluded that their data reveal a KPZ growth dynamics for cell-colony growth, their works do not exhibit a full width function or structure factor collapse~\cite{Huergo2010,Huergo2011,Huergo2012}. 
%
\newpage

\section{Tables}

\begin{table}[h!]
  \begin{center}
    \caption{Each cell's mechanical properties are determined by the values indicated here. They are usually subtly altered to represent different types of cells. The mechanical properties of cells addressed in this work are listed in the table. The symbol $\dagger$ indicates units of $\Delta t$ and $\ast$ units of mean time to cell division, which varies between cell types and is set to 1.0 in \textit{CellSim3D}. For more details and their derivation, please see~\cite{Madhikar2018,Mkrtchyan2014}}
    \label{tab:table1}
	\resizebox{1\columnwidth}{!}
	{\tiny \begin{tabular}{l c c c c}
      \hline
      Parameter & Notation & Sim. Units & \multicolumn{2}{c}{SI Units} \\
      \hline
      Nodes per cell  & $N_\mathrm{c}$ & 180 & - & \\
	  Node mass  & m & 0.04 & 40 & fg\\
	  Bond stiffness  & $k^\mathrm{B}$ & 1000 & 100 & $ \mathrm{nN} \over \mu \mathrm{m}$\\
	  Bond damping coefficient  & $\gamma_\mathrm{int}$ & 100 & 0.01 & $\mathrm{g}\over \mathrm{s}$\\
	  Minimum pressure  & $(PS)_{0}$ & 50 & 0.5 & $\frac{\mathrm{nN}}{\mu \mathrm{m}^{2}}$\\
	  Maximum pressure  & $(PS)_\mathrm{\infty}$ & 65 & 0.65 & $\frac{\mathrm{nN}}{\mu \mathrm{m}^{2}}$\\
	  Pressure growth rate  & $\Delta(PS)$ & 0.002 & $2.0 \times 10^{-5}$ & $\frac{\mathrm{nN}}{\mu \mathrm{m}^{2}}$\\
	  Attraction stiffness  & $K^\mathrm{A}$ & 10-2000 & 1-200 & $\frac{\mathrm{nN}}{\mu \mathrm{m}}$\\
	  Attraction range  & $R^\mathrm{A}_{0}$ & 0.3 & 3 & $\mu m$\\
	  Repulsion stiffness  & $K^\mathrm{R}$ & $ 10 \times 10^{5}$ & $10 \times 10^{4}$ & $\frac{\mathrm{nN}}{m}$ \\
	  Repulsion range  & $R^\mathrm{A}_{0}$ & 0.2 & 2 & $\mu \mathrm{m}$\\
	  Growth count interval  & - & 1000 & $\dagger$ & \\
	  Inter-membrane friction  & $\gamma_\mathrm{ext}$ & 1 & 10 & $\frac{\mu g}{s}$\\
	  Medium friction  & $\gamma_\mathrm{m}$ & 0.4 & 4 & $\frac{\mu \mathrm{g}}{s}$\\
	  Time step  & $\Delta t$ & $1.0 \times 10^{-4}$ & $\ast$ & \\
	  Threshold division volume  & $V^\mathrm{div}$ & 2.9 & 2900 & $\mu \mathrm{m}^{3}$\\
    \end{tabular}}
  \end{center}
\end{table}

\newpage

\section{Figures}

\begin{figure}[h!]
\resizebox{0.4\columnwidth}{!}{ \includegraphics{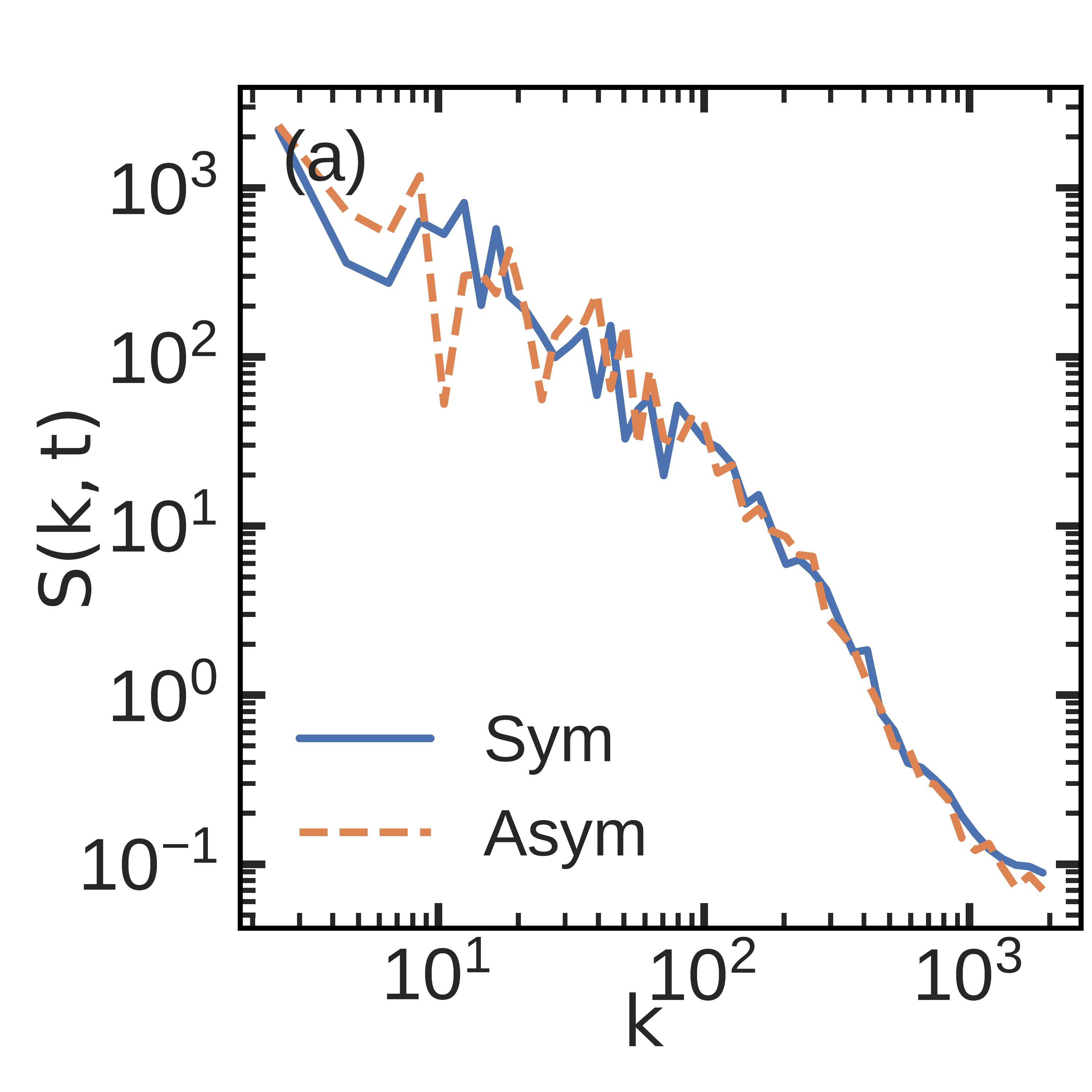} }
\resizebox{0.4\columnwidth}{!}{ \includegraphics{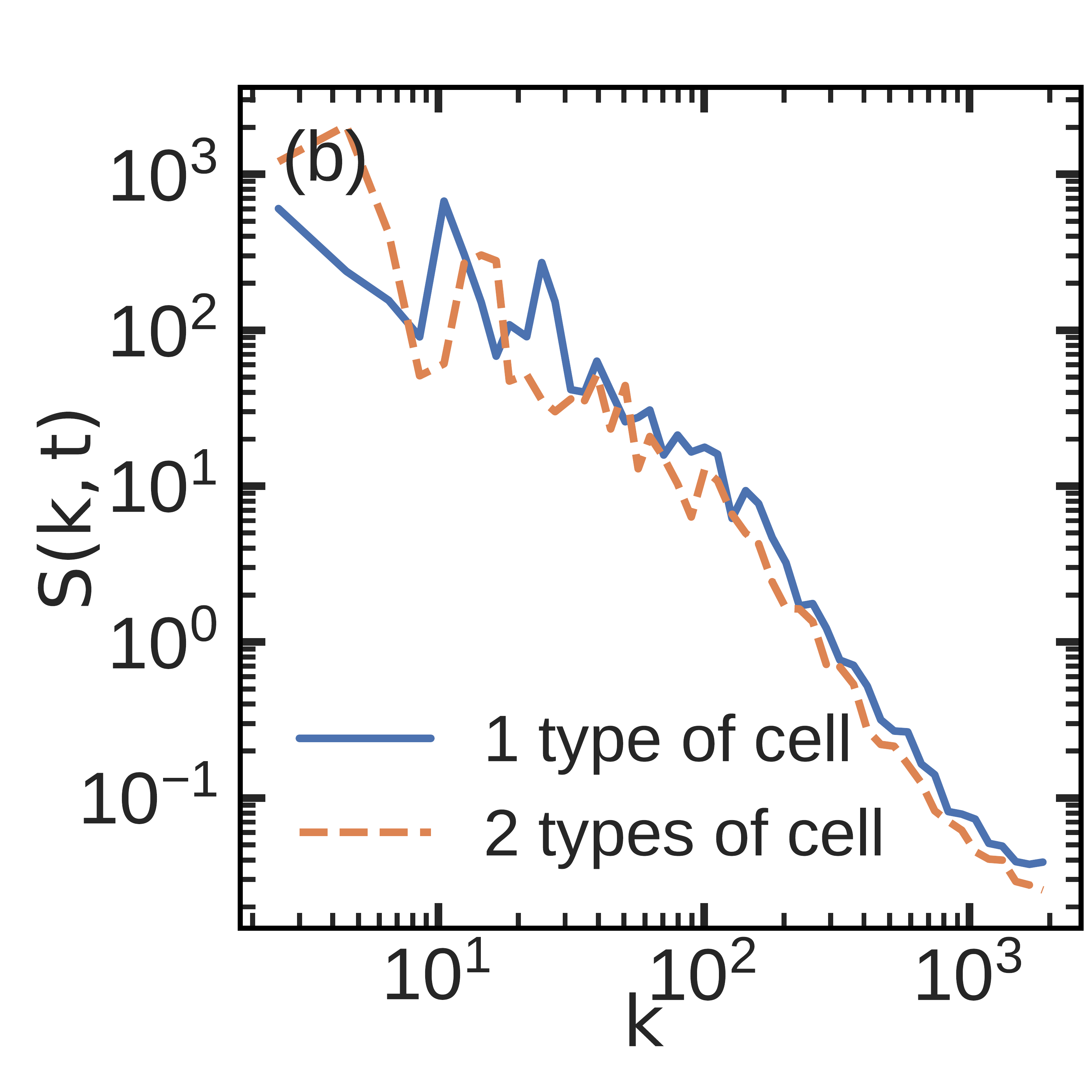} }
\resizebox{0.4\columnwidth}{!}{ \includegraphics{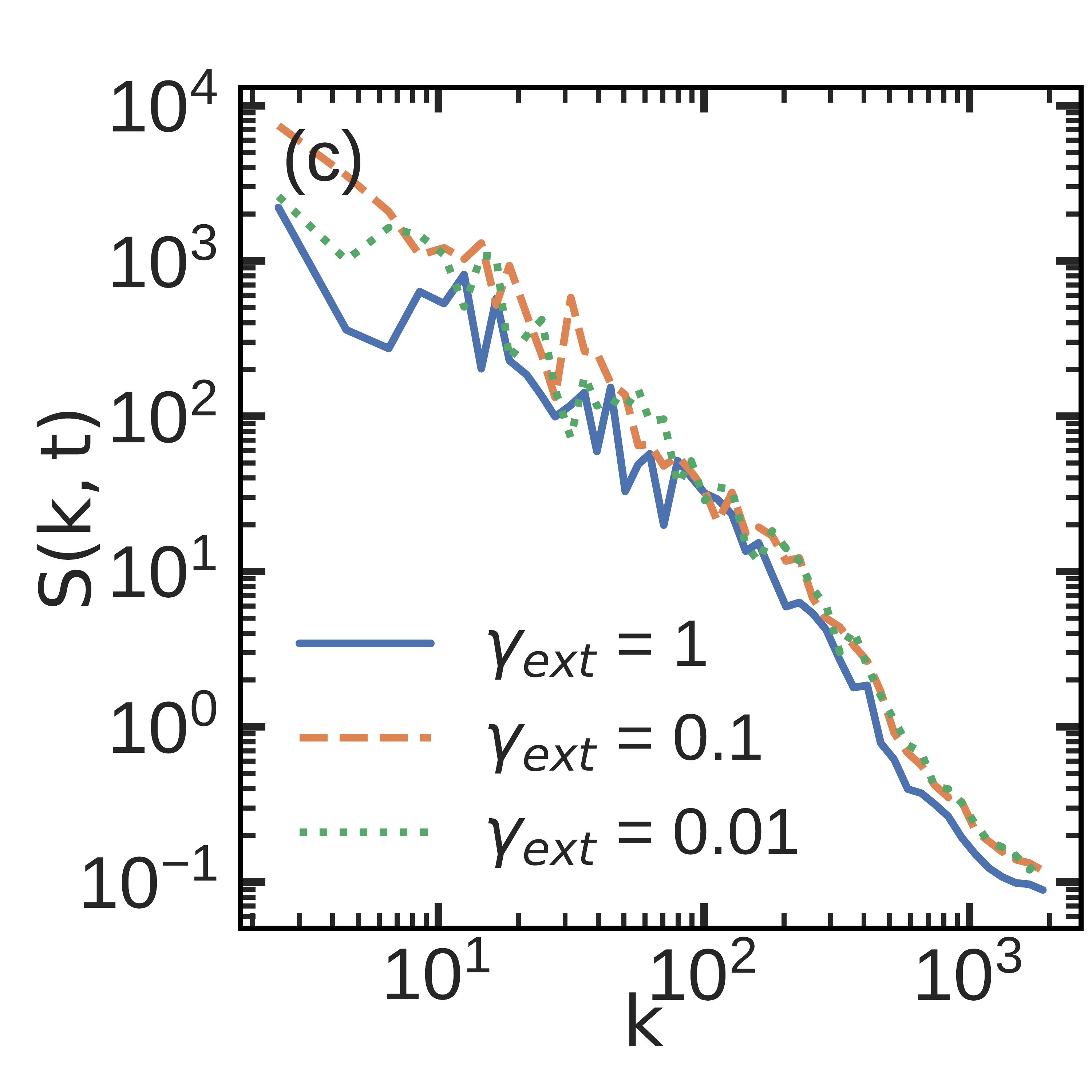} }
\resizebox{0.4\columnwidth}{!}{ \includegraphics{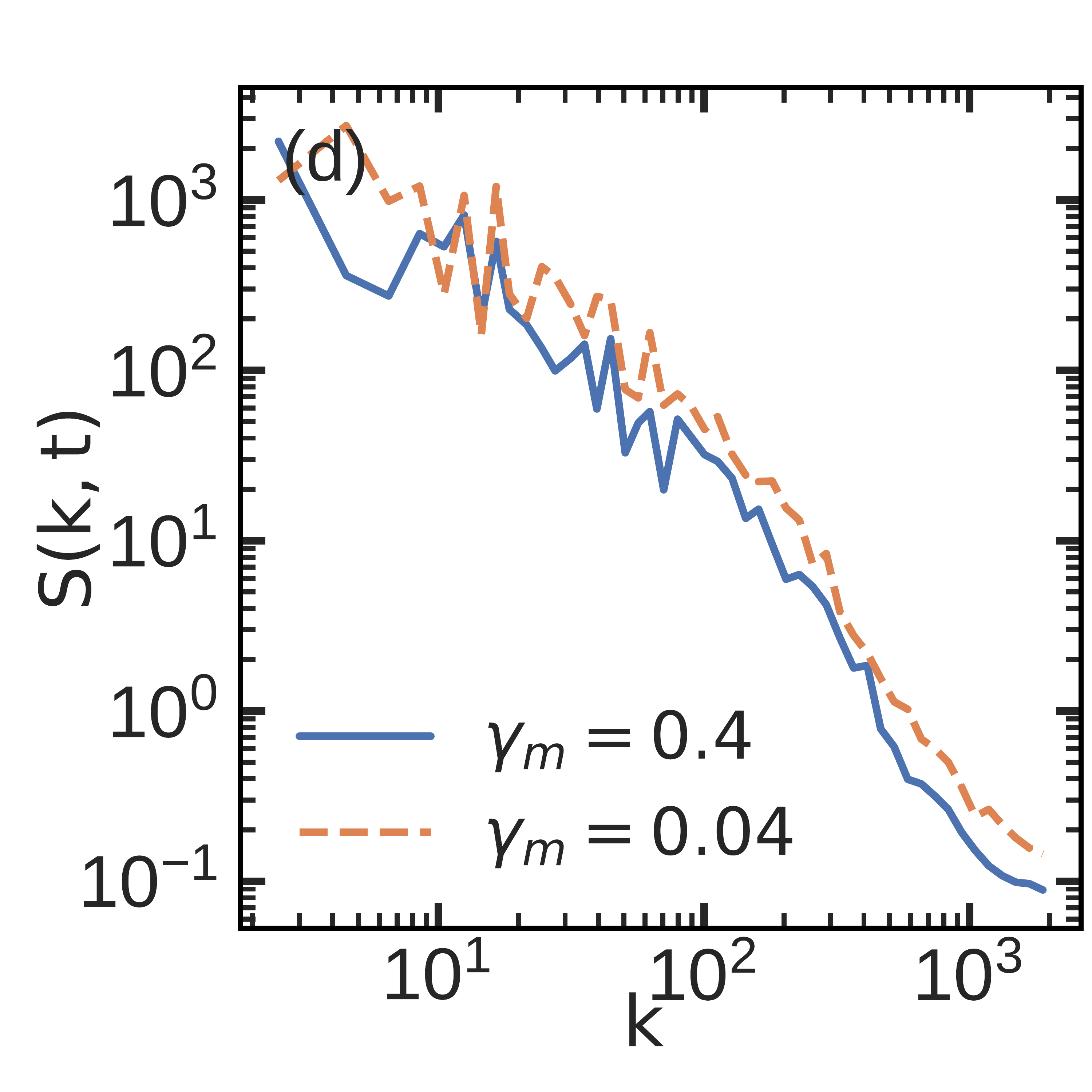} }
\caption{ Log-Log plot of the structure factor of radially expanding interfaces for (a) The division plane is sampled from a unit circle in the epithelial plane, and can yield two symmetric or asymmetric daughter cells in size.  (b) The cell population can contain only one type of cell or two types of cells with different stiffness interacting with each others. (c) different friction coefficients for cell-cell interaction, and (d) different medium friction forces. For units, see Table~\ref{tab:table1}.}
\label{fig:fig7}
\end{figure}

\begin{figure}[h!]
\centering
\includegraphics[height=2.5in]{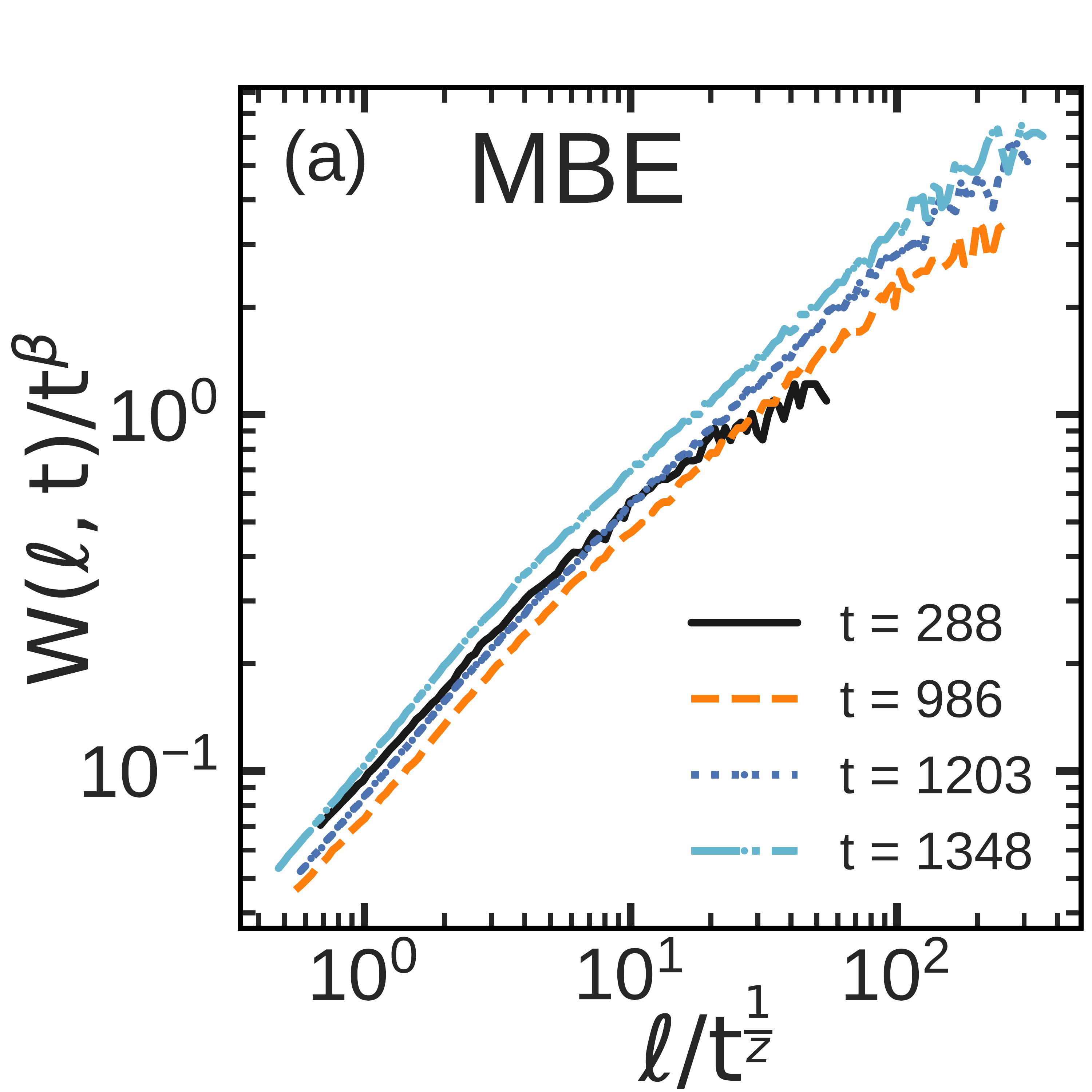}
\includegraphics[height=2.5in]{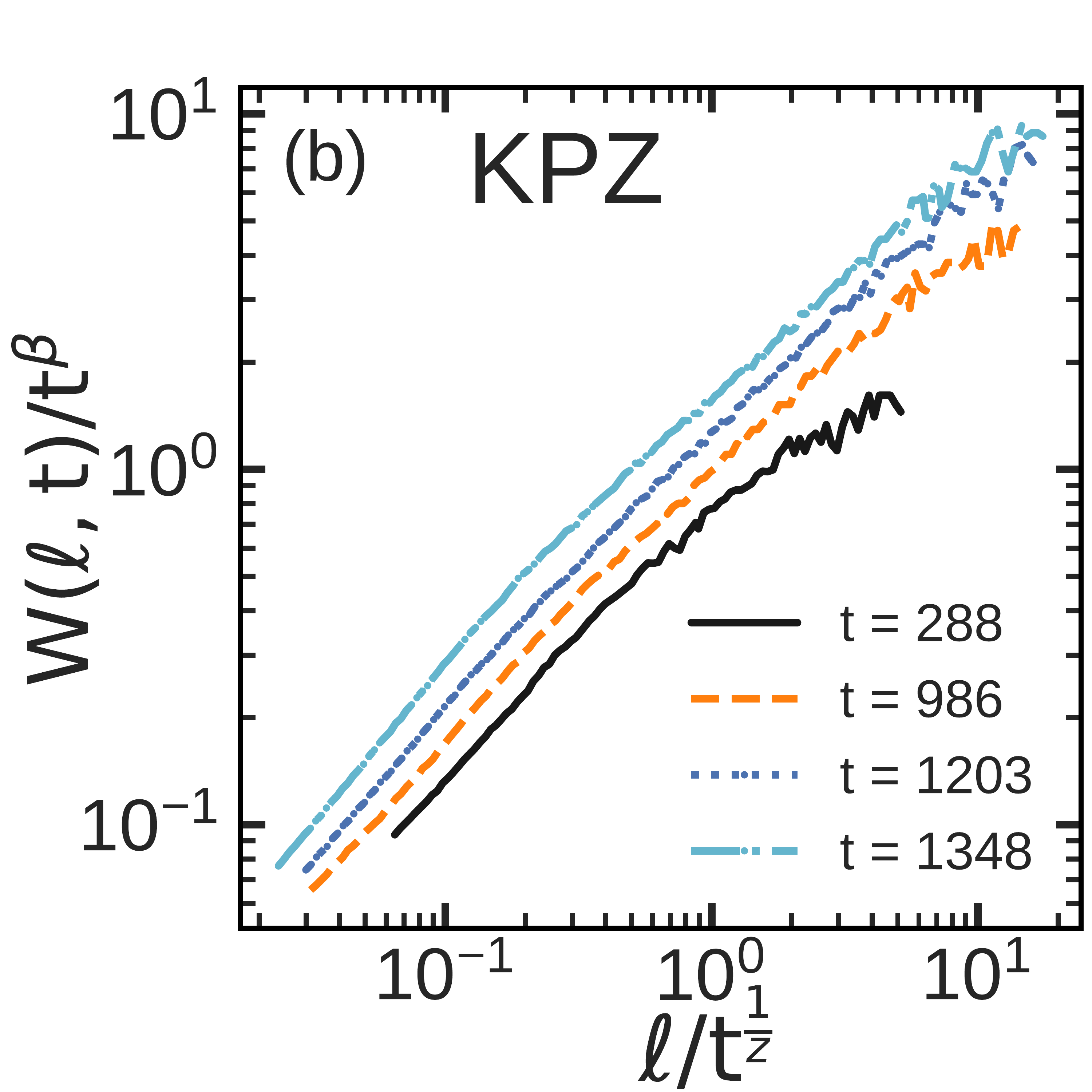}
\caption{ (a) Data collapse using Family-Vicsek relation for width functions digitized from Fig.~3 from Br\'u \textit{et al.}~\cite{Bru2003}, at different times with (a) MBE exponents, $\beta = \frac{3}{8}$ and $z = 4$, and (b) KPZ exponents, $\beta = \frac{1}{3}$ and $z = \frac{3}{2}$. }
\label{fig:fig9}
\end{figure}


%